\documentclass[a4paper,10pt]{revtex4}
\usepackage{mathrsfs}
\usepackage{graphicx}
\usepackage{latexsym}
\usepackage{amsmath}
\usepackage{amssymb}
\usepackage{textcomp}
\usepackage{amsbsy}
\usepackage{graphics}
\usepackage{epstopdf}
\usepackage{color}

\begin{document}

\tolerance=5000

\title{From a Bounce to the Dark Energy Era with $F(R)$ Gravity}
\author{S.~D.~Odintsov,$^{1,2}$\,\thanks{odintsov@ieec.uab.es}V.K.~Oikonomou,$^{3,4,5}$\,\thanks{v.k.oikonomou1979@gmail.com}
Tanmoy~Paul$^{6,7}$\thanks{pul.tnmy9@gmail.com}}
\affiliation{ $^{1)}$ ICREA, Passeig Luis Companys, 23, 08010 Barcelona, Spain\\
$^{2)}$ Institute of Space Sciences (IEEC-CSIC) C. Can Magrans s/n,08193 Barcelona, Spain\\
$^{3)}$ Department of Physics, Aristotle University of
Thessaloniki, Thessaloniki 54124,
Greece\\
$^{4)}$ Laboratory for Theoretical Cosmology, Tomsk State
University of Control Systems and Radioelectronics, 634050 Tomsk,
Russia (TUSUR)\\
$^{5)}$ Tomsk State Pedagogical University, 634061 Tomsk,
Russia\\
$^{6)}$ Department of Physics, Chandernagore College, Hooghly - 712 136, India.\\
$^{(7)}$ Department of Theoretical Physics,\\
Indian Association for the Cultivation of Science,\\
2A $\&$ 2B Raja S.C. Mullick Road,\\
Kolkata - 700 032, India }


\tolerance=5000

\begin{abstract}
In this work we consider a cosmological scenario in which
the Universe contracts initially having a bouncing-like behavior,
and accordingly after it bounces off, it decelerates following a
matter dominated like evolution and at very large positive times
it undergoes through an accelerating stage. Our aim is to study such evolution in the
context of $F(R)$ gravity theory, and confront quantitatively the model
with the recent observations. Using several reconstruction
techniques, we analytically obtain the form of $F(R)$ gravity in two extreme stages of the universe, 
particularly near the bounce and at the late time era respectively. 
With such analytic results and in addition
by employing appropriate boundary conditions, 
we numerically solve the $F(R)$ gravitational equation 
to determine the form of the $F(R)$ for a
wide range of values of the cosmic time. The numerically solved
$F(R)$ gravity realizes an unification of certain cosmological 
epochs of the universe, in particular, from a non-singular bounce 
to a matter dominated epoch and from the matter dominated to a 
late time dark energy epoch. Correspondingly, the
Hubble parameter and the effective equation of state parameter of
the Universe are found and several qualitative features of the
model are discussed. The Hubble radius goes to zero asymptotically
in both sides of the bounce, which leads to the generation of
the primordial curvature perturbation modes near the
bouncing point, because at that time, the Hubble radius diverges
and the relevant perturbation modes are in sub-Hubble scales.
Correspondingly, we calculate the scalar and tensor perturbations
power spectra near the bouncing point, and accordingly we
determine the observable quantities like the spectral index of the
scalar curvature perturbations, the tensor-to-scalar ratio, and as a result, 
we directly confront the present model with the latest Planck
observations. Furthermore the $F(R)$ gravity dark energy
epoch is confronted with the Sne-Ia+BAO+H(z)+CMB data.
\end{abstract}


\maketitle
\section{Introduction}

A major challenge in modern theoretical cosmology is to reveal
whether the Universe emerged from an initial singularity or the
Universe started its expansion from a non-singular bounce-like
phase. In other words, whether the Universe evolve according to
the standard Big-Bang cosmology which must end in a singularity
(known as Big-Bang singularity) when extrapolated backwards, or,
the Universe passes through a bouncing stage leading to a
non-singular behavior of the spacetime curvature. The inflationary
description is an appealing early Universe scenario, due to the
fact that it can produce an almost scale invariant power spectrum,
which is consistent with the latest observations. The inflationary
scenario
\cite{guth,Linde:2005ht,Langlois:2004de,Riotto:2002yw,barrow1,barrow2,nb1,Baumann:2009ds}
requires an early period of accelerated expansion in order to
solve the horizon and flatness problems. However, most
inflationary scenarios have inherent connection with an initial
singularity, at least most of the canonical scalar models of
inflation. Moreover, the observations to date do not undoubtedly
confirm that inflation indeed took place. One of the attractive
alternatives to inflation is the bouncing cosmology scenario
\cite{Brandenberger:2012zb,Brandenberger:2016vhg,Battefeld:2014uga,Novello:2008ra,Cai:2014bea,deHaro:2015wda,Lehners:2011kr,Lehners:2008vx,
Cheung:2016wik,Cai:2016hea,Cattoen:2005dx,Li:2014era,Brizuela:2009nk,Cai:2013kja,Quintin:2014oea,Cai:2013vm,Poplawski:2011jz,
Koehn:2015vvy,Odintsov:2015zza,Nojiri:2016ygo,Oikonomou:2015qha,Odintsov:2015ynk,Koehn:2013upa,Battarra:2014kga,Martin:2001ue,Khoury:2001wf,
Buchbinder:2007ad,Brown:2004cs,Hackworth:2004xb,Nojiri:2006ww,Johnson:2011aa,Peter:2002cn,Gasperini:2003pb,Creminelli:2004jg,Lehners:2015mra,
Mielczarek:2010ga,Lehners:2013cka,Cai:2014xxa,Cai:2007qw,Cai:2010zma,Avelino:2012ue,Barrow:2004ad,Haro:2015zda,Elizalde:2014uba,Das:2017jrl}.
Apart from generating an observationally compatible power spectrum
in some bouncing cosmology scenarios, bouncing cosmology has the
additional advantage that it leads to a singular free evolution
of the Universe. However, the Big-Bang singularity may be a
manifestation for the failure of classical gravity theory which
may be unable to describe the evolution of the Universe at such
small scales. Quite possibly, the quantum generalization of the
gravity theory may resolve the singularity issue; just like the
classical electrodynamics fails to remove the Coulomb potential
singularity at the origin, which, however is resolved by the
quantum electrodynamics. However, in the absence of a fully
accepted quantum gravity theory, bouncing cosmology is the most promising one to deal with the initial singularity issue.\\
Among the various bouncing models proposed so far, the matter
bounce scenario
\cite{deHaro:2015wda,Cai:2008qw,Finelli:2001sr,Quintin:2014oea,Cai:2011ci,
Haro:2015zta,Cai:2011zx,Cai:2013kja,
Haro:2014wha,Brandenberger:2009yt,deHaro:2014kxa,Odintsov:2014gea,
Qiu:2010ch,Oikonomou:2014jua,Bamba:2012ka,deHaro:2012xj,Nojiri:2019lqw,Elizalde:2019tee,Elizalde:2020zcb,
WilsonEwing:2012pu} has attracted a lot of attention since it
produces a scale invariant power spectrum and moreover in a matter
bounce scenario (MBS), the Universe passes through a matter
dominated epoch at the late-time. The MBS is characterized by a
symmetric scale factor where the Hubble radius diverges to
infinity at late times both in the contraction and expanding eras
of the bounce. This indicates that the perturbation modes are
generated far away from the bouncing point, deeply inside the
Hubble radius. Despite the aforementioned successes that the MBS is able to produce a scale invariant power spectrum, the matter bounce
model has some serious drawbacks: (1) in scalar-tensor MBS model,
the scalar power spectrum becomes $exactly$ scale invariant leading
to the spectral index for curvature perturbations as unity, 
which is, in fact, not compatible with the Planck
observations. Such inconsistency was also confirmed in
\cite{Odintsov:2014gea,Nojiri:2019lqw} from a different point of view, in
particular in the context of $F(R)$ gravity theory. However, it is
well known that a $F(R)$ model can be equivalently mapped to a
scalar-tensor model by a suitable conformal transformation of the
spacetime metric and thus the inconsistencies of the scalar power
spectrum (with regard to the Planck observational data) 
in matter bounce scenario for both the scalar-tensor and F(R) theories 
are well justified. (2) The running of the scalar spectral index is
constrained to lie within $-0.0085 \pm 0.0073$ according to the
Planck 2018 observations. However, for the MBS model with a single
scalar field, the running of the spectral index vanishes and thus it 
is not compatible with the observations. This is a direct
consequence of the first problem because if the spectral index is
independent of the perturbation momentum, then the running index
will obviously become zero. (3) The Hubble radius, defined by $r_h
= \frac{1}{aH}$, in the matter bounce scenario monotonically
increases with the cosmic time in the late era and asymptotically
diverges to infinity, due to the reason that the scale factor far
away from the bounce behaves as $\sim t^{2/3}$. Such increasing
behavior of the Hubble radius confirms a decelerating evolution of
the Universe at late times. On other hand, the supernovae
observations
\cite{Perlmutter:1996ds,Perlmutter:1998np,Riess:1998cb} confirm
that the present Universe is dominated by some negative pressure
matter component, known as dark energy, which generates the
accelerated expansion. Thus the dark energy
dominated epoch, which is responsible for the late time acceleration, is not well described by the matter bounce scenario.

It is shown in the literature, that in the so-called quasi-matter
bounce scenario, where the scale factor evolves as
$t^{\frac{2}{3(1+w)}}$ (with $w \neq 0$, note for $w = 0$, it
becomes similar to the exact MBS), it is possible to recover the
consistency of the spectral index and of the running index even in
a single scalar field model \cite{deHaro:2015wda}. However, the
tensor-to-scalar ratio is found to be problematic in the case of a
quasi-matter bounce model; actually the tensor and the scalar
perturbation amplitudes in the quasi-matter bounce scenario are found to be 
comparable to each other leading to the tensor-to-scalar ratio as order of unity. 
On other hand, modified gravity theories
\cite{Nojiri:2017ncd,Nojiri:2010wj,Nojiri:2006ri,Capozziello:2011et,Capozziello:2010zz,delaCruzDombriz:2012xy,Olmo:2011uz,Nojiri:2020wmh}
may provide successful descriptions for both the primordial and
the late-time era of our Universe. In most modified gravity
descriptions of bouncing cosmology, the Hubble radius
monotonically increases at late times and thus the problem to
describe the dark energy epoch of the Universe still persists,
because an increasing behavior of the Hubble radius reveals a
decelerating phase of the universe. Motivated by this problem, in this paper, we
intend to generalize the bouncing scenario which is also compatible with the late-time acceleration as well, in the
context of $F(R)$ gravity theory. It is clear that in order to get a dark energy
dominated phase, the Hubble radius must decrease with respect to the cosmic time at late times.
This in turn indicates that the primordial curvature perturbation
modes generate near the bounce, in contrast to the usual
matter bounce scenario where the perturbation modes generate
far away from the bounce deeply in the contracting era. Keeping
these issues in mind, we try to provide a cosmological scenario 
which unifies certain cosmological epochs of the universe, in particular, 
from a non-singular bounce to a matter dominated era and from the matter dominated to 
a late time dark energy epoch, in the context of $F(R)$ gravity. In this regard, we would like to
mention that the merging of bounce with the dark energy epoch was
studied earlier \cite{Odintsov:2016tar}, however in a different context, 
where the universe evolution was not considered to be symmetric with respect to the bounce point, unlike to our 
present work where we will consider a symmetric behavior of the scale factor as a function of the cosmic time and 
consequently the Hubble radius asymptotically goes to zero in both sides of the bounce.

The outline of this paper is as follows : after briefly describing
the essential features of $F(R)$ gravity in Sec.[\ref{sec_F(R)}],
we will demonstrate the behavior of $F(R)$ gravity near the bounce
and during the late-era respectively in
Sec.[\ref{sec_reconstruction}] and consequently in
Sec.[\ref{sec_estimation}], we will perform the scalar and tensor
perturbations. Some numerical treatment of several qualitative
aspects of the theory at hand are presented in
Sec.[\ref{sec_unification}]. Finally the conclusions follow in the end of the paper.

\section{Essential features of $F(R)$ gravity}\label{sec_F(R)}

Let us briefly recall some basic features of $F(R)$ gravity, which are necessary for our presentation, for reviews on
this topic see \cite{Nojiri:2010wj,Nojiri:2017ncd,Capozziello:2011et}. The gravitational action of $F(R)$ gravity in vacuum is equal to,
\begin{eqnarray}
 S = \frac{1}{2\kappa^2} \int d^4x \sqrt{-g} F(R)
 \label{basic1}
\end{eqnarray}
where $\kappa^2$ stands for $\kappa^2 = 8\pi G = \frac{1}{M_\mathrm{Pl}^2}$ and also $M_\mathrm{Pl}$ 
is the reduced Planck mass. By using the metric formalism, we vary the
action with respect to the metric tensor $g_{\mu\nu}$, and the gravitational equations read,
\begin{eqnarray}
 F'(R)R_{\mu\nu} - \frac{1}{2}F(R)g_{\mu\nu} - \nabla_{\mu}\nabla_{\nu}F'(R) + g_{\mu\nu}\Box F'(R) = 0
 \label{basic2}
\end{eqnarray}
where $R_{\mu\nu}$ is the Ricci tensor constructed from $g_{\mu\nu}$. Since the present article is devoted to cosmological context,
the background metric of the Universe will be assumed to be a flat Friedmann-Robertson-Walker (FRW) metric,
\begin{eqnarray}
 ds^2 = -dt^2 + a^2(t)\big[dx^2 + dy^2 + dz^2\big]
 \label{basic3}
\end{eqnarray}
with $a(t)$ being the scale factor of the Universe. For this
metric, the temporal and spatial components of Eq.(\ref{basic2})
become,
\begin{eqnarray}
3H^2&=&-\frac{f(R)}{2} + 3\big(H^2 + \dot{H}\big)f'(R) - 18\big(4H^2\dot{H} + H\ddot{H}\big)f''(R)\nonumber\\
0&=&\frac{F(R)}{2} - \big(3H^2 + \dot{H}\big)F'(R) + 6\big(8H^2\dot{H} + 4\dot{H}^2 + 6H\ddot{H}
+ \dddot{H}\big)F''(R) + 36\big(4H\dot{H} + \ddot{H}\big)^2F'''(R)
\label{basic4}
\end{eqnarray}
respectively, where $H = \dot{a}/a$ is the Hubble parameter of the
Universe and $f(R)$ is the deviation of $F(R)$ gravity from the
Einstein gravity, that is $F(R) = R + f(R)$. Comparing the above
equations with the usual Friedmann equations, it is easy to
understand that $F(R)$ gravity provides a contribution in the
energy-momentum tensor, with its effective energy density
($\rho_{eff}$) and pressure ($p_{eff}$) being given by,
\begin{eqnarray}
 \rho_{eff} = \frac{1}{\kappa^2}\bigg[-\frac{f(R)}{2} + 3\big(H^2 + \dot{H}\big)f'(R) - 18\big(4H^2\dot{H} + H\ddot{H}\big)f''(R)\bigg]
 \label{ed}
\end{eqnarray}
\begin{eqnarray}
 p_{eff} = \frac{1}{\kappa^2}\bigg[\frac{f(R)}{2} - \big(3H^2 + \dot{H}\big)f'(R) + 6\big(8H^2\dot{H} + 4\dot{H}^2 + 6H\ddot{H}
+ \dddot{H}\big)f''(R) + 36\big(4H\dot{H} + \ddot{H}\big)^2f'''(R)\bigg]
\label{pressure}
\end{eqnarray}
respectively. Thus, the effective energy-momentum tensor (EMT)
depends on the form of $F(R)$, as expected. Therefore, different
forms of $F(R)$ will lead to different evolution of the Hubble
parameter. In the present context, we will use such effective EMT
of $F(R)$ gravity to realize the cosmological evolution of the
Universe.


\section{Reconstruction of $F(R)$ gravity: Realization of Bounce and Dark Energy Era}\label{sec_reconstruction}

In this section, we reconstruct the $F(R)$ gravity from
Eq.(\ref{basic4}) by considering a certain form of the scale
factor of the Universe. In particular, we reconstruct the $F(R)$
gravity in the two distinct eras, namely near the bouncing point
and at the late-time epoch, which are the primary subjects in the
following two subsections respectively. Clearly such forms of
$F(R)$ will describe the gravity theory in the two extreme stages
of the Universe and thus will not be able to reveal the Universe
evolution as a whole. However later in
Sec.[\ref{sec_unification}], we will numerically solve
Eq.(\ref{basic4}) and will determine the form of $F(R)$ for a wide
range of cosmic times, which will provide an unification of certain cosmological epochs of the universe, 
particularly from a bounce to matter dominated 
era, followed by a late time accelerating phase. During the numerical solution of
Eq.(\ref{basic4}), the form of $F(R)$ in the two distinct eras,
which will be analytically evaluated in the following two
subsections, will act as boundary conditions.

\subsection{Reconstruction near the bounce}\label{sec_rec_bounce}

In this section, we reconstruct the form of the $F(R)$ gravity
near the bounce of the Universe. The Universe's evolution in a
general bouncing cosmology, consists of two eras, an era of
contraction and an era of expansion. Some of the well known
functional forms of the scale factor which correspond to a
non-singular bounce, have the form, $a(t) = e^{\alpha t^2}$, $a(t)
= \cosh{t}$, $a(t) = (a_0t^2 +1)^n$ and so on. In general, the
non-singular bounce of a scale factor is characterized by
\begin{eqnarray}
 a(t_b) \neq 0~~~~~~~~~~,~~~~~~~~~~\dot{a}(t_b) = 0~~~~~~~~~~~,~~~~~~~~~~~\ddot{a}(t_b) > 0
 \nonumber
\end{eqnarray}
where $t_b$ is the cosmic time when the bounce occurs. The above
conditions lead to a finite Kretschmann scalar ($K =
R_{\mu\nu\alpha\beta}R^{\mu\nu\alpha\beta}$) and thus the Universe
becomes free of the initial singularity (known as Big-Bang
singularity). Keeping these conditions in mind, we consider the
scale factor near the bounce as,
\begin{eqnarray}
 a_b(t) = 1 + \alpha t^2
 \label{rec_bounce scale}
\end{eqnarray}
(the suffix 'b' stands for near $bounce$ scale factor) with
$\alpha$ being a free parameter having mass dimension [+2] and the
bounce happens at $t = 0$. The above form of the scale factor can
be thought as a Taylor series expansion of $a(t)$ around $t = 0$ and keeping
up-to quadratic order in cosmic time (t). We neglect the higher orders of $t$ in the Taylor
expansion of $a(t)$ as, in this section, we are interested to
reconstruct the $F(R)$ gravity near the bouncing point. 
The linear order of $t$ in the Taylor expansion
vanishes due to the condition $\dot{a} = 0$, necessary for a
bounce. Moreover the condition $\ddot{a}(0) > 0$ indicates that
the parameter $\alpha$ must be positive and thus we take $\alpha >
0$ in the present context. Here we would like to mention that in the next sections, we
will also provide a complete form of the scale factor which is valid for a wide range of cosmic time and 
may yield the behavior (\ref{rec_bounce scale}) near the
bouncing point. The scale factor of Eq.(\ref{rec_bounce
scale}) immediately leads to the following Hubble parameter ($H =
\frac{\dot{a}}{a}$) and Ricci scalar ($R(t)$) as,
\begin{eqnarray}
 H(t)&=&\frac{2\alpha t}{1 + \alpha t^2} \simeq 2\alpha t~~~,\nonumber\\
 R(t)&=&12H^2 + 6\dot{H} = \frac{12\alpha(1 + 3\alpha t^2)}{(1 + \alpha t^2)^2} \simeq 12\alpha + 12\alpha^2t^2
 \label{rec_bounce ricci}
 \end{eqnarray}
 respectively, with the $H(t)$ and $R(t)$ being considered up to $\mathcal{O}(t^2)$, similar to the case of the scale factor. However
 Eq.(\ref{rec_bounce ricci}) clearly indicates that the Hubble parameter varies linearly with $t$ and goes to zero at the bouncing point, while
 the Ricci scalar, on the other hand, becomes $R(0) = 12\alpha$. Later, during the calculations of scalar and tensor perturbations,
 we will show that the parameter $\alpha$ should be at the order $\sim 10^{-8}/\kappa^2$ to make the model compatible with the Planck constraints
 and thus the Ricci scalar becomes $\sim 10^{28}\mathrm{GeV}^2$ (with $\frac{1}{\kappa^2} = M_{\mathrm{Pl}}^2 = 10^{36}\mathrm{GeV}^2$)
 at the bouncing point. In order to reconstruct the $F(R)$ gravity,
 we will use Eq.(\ref{basic4}), for which we determine the following quantities (appearing in Eq.(\ref{basic4})) with the scale
 factor given in Eq.(\ref{rec_bounce scale}),
 \begin{eqnarray}
 H^2 + \dot{H}&=&\frac{2\alpha}{1 + \alpha t^2} \simeq 2\alpha - 2\alpha^2t^2~~~,\nonumber\\
 4H^2\dot{H} + H\ddot{H}&=&\frac{8\alpha^3t^2(1 - 3\alpha t^2)}{(1 + \alpha t^2)^4} \simeq 8\alpha^3t^2
 \nonumber
 \end{eqnarray}
 With the above expressions, Eq.(\ref{basic4}) becomes,
 \begin{eqnarray}
  24\alpha(R - 12\alpha)f_b''(R) + (R - 24\alpha)f_b'(R) + f_b(R) + 2(R - 12\alpha) = 0\label{rec_bounce eom}
 \end{eqnarray}
Solving the above equation for $f(R)$, we get,
\begin{eqnarray}
  f_b(R) = \bigg(\frac{12\alpha D}{\sqrt{e}} - 1\bigg)R - D\sqrt{6\alpha\pi}~e^{-\frac{R}{24\alpha}}\big(R - 12\alpha\big)^{3/2}~
  Erfi\bigg[\frac{\sqrt{R - 12\alpha}}{2\sqrt{6\alpha}}\bigg]
  \label{rec_bounce sol1}
 \end{eqnarray}
where $Erfi[z]$ is the imaginary error function defined as $Erfi[z] = -iErf[iz]$ with $Erf[z]$ being the error function and '$i$' is the imaginary unit.
Moreover $D$ is an integration constant having mass dimension [-2]. The above solution of $f_b(R)$ immediately leads to the form
of $F_b(R) = R + f_b(R)$ as,
\begin{eqnarray}
  F_b(R) = \frac{12\alpha D}{\sqrt{e}}R - D\sqrt{6\alpha\pi}~e^{-\frac{R}{24\alpha}}\big(R - 12\alpha\big)^{3/2}~
  Erfi\bigg[\frac{\sqrt{R - 12\alpha}}{2\sqrt{6\alpha}}\bigg]
  \label{rec_bounce sol2}
 \end{eqnarray}
 The presence of $e^{-\frac{R}{24\alpha}}$ and $Erfi\bigg[\frac{\sqrt{R - 12\alpha}}{2\sqrt{6\alpha}}\bigg]$ in the
 right hand side of Eq.(\ref{rec_bounce sol2}) confirm that the $F_b(R)$ contains
 all the positive integer powers of $R$ where the coefficient of the linear Ricci scalar is other than unity. This indicates
 a deviation from Einstein gravity. However recall, the form of $F(R)$ in Eq.(\ref{rec_bounce sol2}) represents the
 gravity theory near the bounce and in a later Sec.[\ref{sec_unification}],
 during the numerical reconstruction of $F(R)$ gravity for a wide range of cosmic
 times, we will show that the late time $F(R)$ indeed matches with the Einstein gravity.

 \subsection{Reconstruction in the late time epoch}\label{sec_rec_late}

In this section, we apply a reconstruction scheme to describe
the behavior of the Universe at late times. Recall, in the
previous section during the reconstruction near the bounce, we
considered a scale factor (suitable for bounce) and then, by
solving Eq.(\ref{basic4}), we determine the corresponding form of
$F(R)$. However the late-time reconstruction method which we are
going to apply in this section, is slightly different in comparison to that of the 
earlier one, in particular, we will start with a form of $F(R)$ (rather than a scale factor)
suitable for a dark energy model and then reconstruct the corresponding Hubble parameter 
from the gravitational equation of motion.

\begin{figure}[!h]
\begin{center}
 \centering
 \includegraphics[width=18pc]{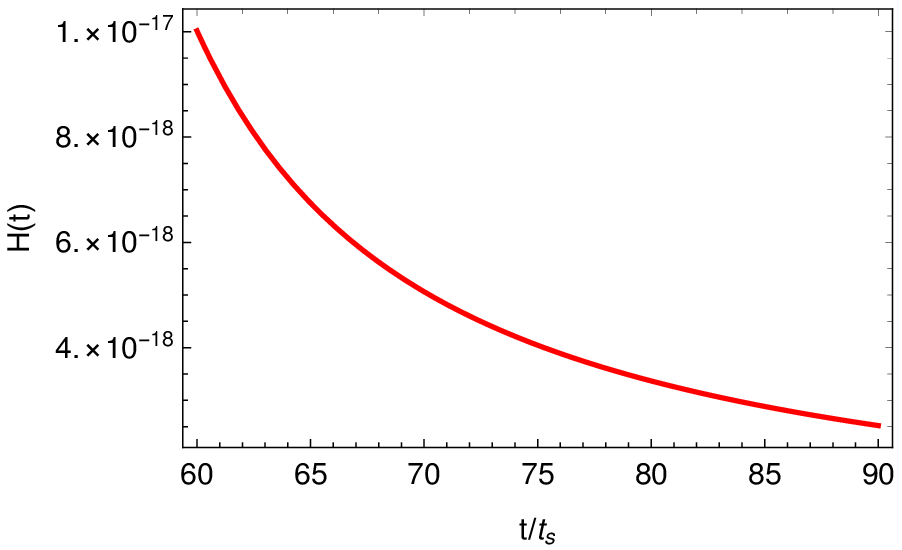}
 \includegraphics[width=18pc]{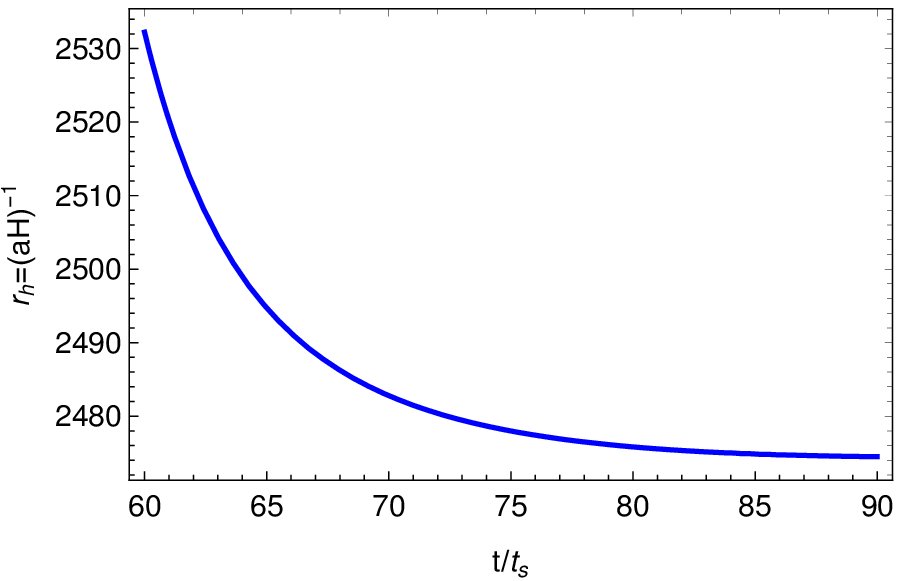}
 \caption{$Left~plot$ : $H(t)$ (along y-axis) vs. $\frac{t}{t_s}$ (along x-axis) for $6\times10^{17} \leq t \leq 9\times10^{17}\mathrm{sec.}$ 
 The Hubble parameter
 along the y-axis is in the unit of sec$^{-1}$. $Right~plot$ : $r_h^{(l)}(t)$ (along y-axis) vs. $\frac{t}{t_s}$ for
 $6\times10^{17} \leq t \leq 9\times10^{17}$ sec, where the Hubble radius is in the unit of sec. The decreasing behavior of $r_h^{(l)}(t)$
 clearly indicates an accelerating stage of the Universe during late time.}
 \label{plot_rec_late}
\end{center}
\end{figure}

In order to reproduce the current acceleration of the Universe,
several versions of viable modified gravity including the
so-called one-step models have been proposed in Ref.
\cite{Hu:2007nk,Appleby:2007vb,Cognola:2007zu,Elizalde:2010ts,Linder:2009jz}.
The simplest one is given by the so-called exponential gravity
which includes an exponential function of the Ricci scalar in the
action. We consider such exponential $F(R)$ model in the present
section, which is known to provide a good dark energy model as
described in \cite{Odintsov:2018qug,Odintsov:2017qif}. In
particular, we consider,
\begin{eqnarray}
 F_l(R) = R - 2\Lambda\bigg(1 - e^{-\frac{\beta R}{2\Lambda}}\bigg)
 \label{rec_late_FR}
\end{eqnarray}
 with $\beta$ and $\Lambda$ are two model parameters
 having mass dimensions [0] and [+2] respectively and the suffix 'l' is for ``late'' time epoch.
 The exponential correction over the usual Einstein-Hilbert action
 becomes important at cosmological scales and at late-times, providing an alternative to the dark energy problem. Due to to the Supernovae
 Ia (Sne-Ia) \cite{Suzuki:2011hu,Scolnic:2017caz}, Baryon Acoustic Oscillations (BAO) \cite{Eisenstein:2005su,bao,Delubac:2014aqe,Wang:2016wjr},
 Cosmic Microwave Background (CMB) \cite{Ade:2013zuv,Ade:2015xua,Wang:2013mha,Huang:2015vpa}
 and $H(z)$ \cite{Simon:2004tf,Moresco:2016mzx,Moresco:2015cya,Ratsimbazafy:2017vga} data, the parameters
 $\beta$ and $\Lambda$ are well constrained, particularly the $F(R)$ model (\ref{rec_late_FR}) is best fitted with
 Sne-Ia+BAO+H(z)+CMB data for the parametric regimes given by : $\beta = 3.98^{+\infty}_{-2.46}$ and $\Lambda = 1.2\times10^{-84}\mathrm{GeV}^2$
 \cite{Odintsov:2018qug}. Such parametric ranges also make the model compatible with Sne-Ia+BAO+H(z) data. In effect,
 we stick with $\beta = 4$ and $\Lambda = 1.2\times10^{-84}\mathrm{GeV}^2$ throughout the paper. With these viable ranges of the model
 parameters along with the $F(R)$ of Eq.(\ref{rec_late_FR}),
 now we are going to solve the gravitational equation of motion to reconstruct the evolution of the Hubble parameter during the dark energy
 epoch. Plugging the above form of $F(R)$ into Eq.(\ref{basic4}), we
 get,
 \begin{eqnarray}
  -3H_l^2 + \Lambda\big(1 - e^{-\frac{\beta R}{2\Lambda}}\big) -3\beta e^{-\frac{\beta R}{2\Lambda}}\big(H_l^2 + \dot{H}_l\big)
  - \frac{9\beta^2}{\Lambda} e^{-\frac{\beta R}{2\Lambda}} \big(4H_l^2\dot{H}_l + H_l\ddot{H}_l\big) = 0
  \label{rec_late_eq1}
 \end{eqnarray}
 where the ``dot'' represents $\frac{d}{dt}$. However getting an analytic solution of Eq.(\ref{rec_late_eq1}) is
troublesome, so we proceed to solve it numerically and for this
purpose, we introduce a dimensionless time scale ($t_r$) and a
dimensionless Hubble parameter ($H_r$) defined as follows:
 \begin{eqnarray}
  t_r = \frac{t}{t_s}~~~~~~~~~~~~~~~~~,~~~~~~~~~~~~~~~~~~H_r = \frac{1}{a_l}\frac{da_l}{dt_r}
  \nonumber
 \end{eqnarray}
with $t_s = 10^{16}$ sec i.e at the order of the
present age of the Universe. The actual Hubble parameter
($H_l(t)$) is related with the rescaled one as $H_l(t) =
\frac{1}{t_s}H_r(t_r) = 10^{-16}H_r$ (sec)$^{-1}$, moreover the
derivative of the Hubble parameter in the two time coordinates are
connected as $\dot{H}_l(t) = \frac{1}{t_s^2}\frac{dH_r}{dt_r}$
(note, $\frac{d}{dt}$ is represented by a ``dot'' while the
$\frac{d}{dt_r}$ is mentioned by as it is). In view of these
relations, Eq.(\ref{rec_late_eq1}) can be expressed in terms of
the rescaled quantities as follows,
\begin{eqnarray}
 -3H_r^2 + \Lambda t_s^2\big(1 - e^{-\frac{\beta R}{2\Lambda}}\big) -3\beta e^{-\frac{\beta R}{2\Lambda}}\big(H_r^2 + \frac{dH_r}{dt_r}\big)
  - \frac{9\beta^2}{\Lambda t_s^2} e^{-\frac{\beta R}{2\Lambda}} \big(4H_r^2\frac{dH_r}{dt_r} + H_r\frac{d^2H_r}{dt^2}\big) = 0
  \label{rec_late_eq2}
\end{eqnarray}
where $\frac{R}{\Lambda} = \frac{1}{\Lambda t_s^2}\big(12H_r^2 + 6
\frac{dH_r}{dt_r}\big)$. With $\beta = 4$ and $\Lambda t_s^2 =
1.52^2\times10^{-4}$ (the conversion $1$ sec. $=
1.52\times10^{24}\mathrm{GeV}^{-1}$ may be useful),
Eq.(\ref{rec_late_eq2}) is solved numerically and is given in the
left part of Fig.[\ref{plot_rec_late}], where the Hubble parameter
is plotted for $60 \leq t_r \leq 90$ (or equivalently
$6\times10^{17} \leq t \leq 9\times10^{17}\mathrm{sec.}$) during the late
era. Moreover we use $H_r(t_r = 60) = 0.1$ and
$\frac{dH_r}{dt_r}\bigg|_{t_r = 60} = -0.1$ as the boundary
conditions to solve the above equation numerically. The x-axis of
the left part of Fig.[\ref{plot_rec_late}] is the rescaled time
$t_r$ and by using the relation $H_l = 10^{-16}H_r\mathrm{sec}^{-1}$,
we give the actual Hubble parameter $H_l(t)$ (in the unit of
(sec)$^{-1}$) along the y-axis. The solution of the the Hubble
parameter immediately leads to the evolution of the Hubble radius
($r_h^{(l)} = \frac{1}{H_l\exp{\big[\int H_l(t)~dt\big]}}$) as
shown in the right part of Fig.[\ref{plot_rec_late}] where
once again, the x-axis is the rescaled time coordinate and the y-axis represents the actual Hubble radius i.e $r_h^{(l)}(t) = \big(a_l(t)H_l(t)\big)^{-1}$.\\
Fig.[\ref{plot_rec_late}] clearly reveals that the Hubble radius
decreases with time and leading to an accelerating stage of the
Universe at late time. Therefore as a whole, the $F(R)$
(\ref{rec_late_FR}) successfully provides a viable dark energy
model in respect to Sne-Ia+BAO+H(z)+CMB data where the evolution of
the Hubble parameter and the Hubble radius are given in
Fig.[\ref{plot_rec_late}]. Moreover, here it may be mentioned that the exponential $F(R)$ model
has a Schwarzschild de-Sitter solution in the regime $R \gg
\Lambda \sim 10^{-84}\mathrm{GeV}^2$. For example, in the Solar System regime 
$R^{*} \simeq 10^{-78}\mathrm{GeV}^2$, the exponential $F(R)$ model can be approximated as $F(R^{*}) = R^{*} -
2\Lambda$ (note that $R^{*}$ is greater than
the $\Lambda$) and thus in this case, 
the Schwarzschild de-Sitter solution can be obtained, 
which is indeed a stable solution as depicted by some of our authors in \cite{Elizalde:2010ts}.\\

Therefore Eqs.(\ref{rec_bounce sol2}) and (\ref{rec_late_FR})
represent the $F(R)$ gravity in the two situations, one depicting
the early bounce of the Universe where the scale factor behaves as
$a(t) = 1+\alpha t^2$ ($= a_b(t)$) and the other providing a late
time accelerating phase when the scale factor ($a_l(t)$) or more
explicitly the Hubble parameter $H_l(t)$ goes according to the
Fig.[\ref{plot_rec_late}] (the corresponding scale factor can be
easily obtained by using $a_l(t) = e^{\int H_l(t)dt}$). Till now,
we have described the bounce and the dark energy (DE) era in a
disjoint manner i.e $F_b(R)$ describes the Universe near the
bounce but not the whole cosmological epochs, while the $F_l(R)$
reveals the behavior only at present time. Thus the immediate
question that springs to mind is, what is the form of $F(R)$ for
the entire range of cosmic time, which will further demonstrate
the unification of bounce and dark energy era followed by a
deceleration stage in the intermediate region? As a deceleration
era, we may consider a matter dominated epoch in between the
bounce and the present era. With this consideration, in Sec.[\ref{sec_unification}], we
will reconstruct the full $F(R)$ gravity, however numerically, by
taking the forms of $F_b(R)$ and $F_l(R)$ as boundary
conditions. In the method of numerical reconstruction, we need an
appropriate scale factor (or equivalently Hubble radius)
describing the Universe evolution for a wide range of cosmic time. 
For this purpose, we may consider an ansatz of
the scale factor as follows,
\begin{eqnarray}\label{anszatsunifiedform}
a(t)&=&e^{\alpha t^2}\tanh{\bigg[\beta_1\bigg(1 -
\frac{t}{t_p}\bigg)\bigg(1 - \frac{t}{t_I}\bigg)\bigg]} +
\bigg(\frac{t}{T}\bigg)^{2/3}\tanh{\bigg[\beta_2\bigg(\frac{t}{t_b} - 1\bigg)\bigg(1 - \frac{t}{t_p}\bigg)\bigg]}\nonumber\\
&+&a_l(t)\tanh{\bigg[\beta_3\bigg(\frac{t}{t_b} - 1\bigg)\bigg(\frac{t}{t_I} - 1\bigg)\bigg]}
\end{eqnarray}
where $t_b$ is the bouncing time or even the horizon crossing time, $t_p$ is the present time instance and $t_I$ is an intermediate time. The above
scale factor behaves as $a_b(t) = 1 + \alpha t^2$ near $t \simeq t_b \ll t_I \ll t_p$ i.e near the bounce, however goes as
$a_I(t) = \big(t/T\big)^{2/3}$ near $t_b \ll t = t_I \ll t_p$ and as $a_l(t) = e^{\int H_l(t)dt}$ near the present time i.e
$t \simeq t_p$. The parameters $\beta_i$ can be fixed by considering the corresponding hyperbolic tangent 
function tending to unity at the respective time.\\

Though the scale factor (\ref{anszatsunifiedform}) provides the correct evolution of the universe
near the bounce and at the late stage and also at a certain instance of the intermediate stage ($t \simeq t_I$), but it does not lead to
correct evolution of the universe for the whole intermediate epoch. Thus to address the universe evolution smoothly for a wide range of cosmic time,
we may consider $a(t)$ as,
\begin{eqnarray}
 a(t)&=&1 + \alpha t^2 = a_b(t)~~,~~~~~~~~~~~~~~~~~~~~~~~~~near~the~bounce,~for~~~~~~~t\simeq 0^{+}\nonumber\\
 a(t)&=&\bigg(\frac{t}{T}\bigg)^{2/3} = a_I(t)~~,~~~~~~~in~the~intermediate~regime,~for~~~~~~~0^{+} \lesssim t \lesssim 5\times10^{17}~sec\nonumber\\
 a(t)&=&e^{\int H_l(t)dt} = a_l(t)~~,~~~~~~~~~~~~~~~~~~~~~during~late~time,~for~~~~~~~~t \gtrsim 6\times10^{17}~sec
 \label{scale_ansatz}
\end{eqnarray}
(with $T$ being an arbitrary parameter with dimension $sec$) and
the transition from the bounce to the matter dominated and from
matter dominated to the dark energy era will be obtained by the
method of numerical interpolation. However, the process of
interpolation requires appropriate choices for the values of the
free parameters present in our model. It may be observed that
there are three parameters hanging around in the theory, $\alpha$,
$\Lambda$ and $\beta$. The parameter $\alpha$ appears from the
near-bounce scale factor and thus can be named as ``early stage
parameter'', while $\Lambda$ and $\beta$ appear during the
reconstruction in the late-time and thus the name - ``late stage
parameters''. The late stage parameters have been already
estimated in view of the viability of the exponential $F(R)$ as a
successful dark energy model. However, the early stage parameter
$\alpha$ is still remaining to be determined and in the next
section, we will estimate $\alpha$ from the latest Planck 2018
observational results.

\section{Cosmological perturbation: Estimation of early stage parameters}\label{sec_estimation}

Being the early stage parameter, $\alpha$ can be determined from
various primordial observable quantities like the scalar spectral
index ($n_s$), the tensor-to-scalar ratio ($r$) by directly
confronting their theoretical expectations with the Planck 2018 observations. In this
section, we consider the spacetime fluctuations over the FRW
metric and consequently calculate various observable quantities
like the scalar spectral index and tensor-to-scalar ratio. In a
bouncing Universe, the primordial perturbation modes (relevant to
the present day observation) generate either near the bounce
or at a distant past far away from the bouncing point, depending
upon the late-time behavior of the Hubble radius. Thereby before
moving to the explicit perturbation calculation, it is important
to analyze when the perturbation modes generate in the
present context of bouncing Universe. In all the bouncing models,
the Hubble parameter becomes zero and thus the comoving Hubble
radius, defined by $r_h = \frac{1}{aH}$ (where $H$ is the Hubble
parameter), diverges at the bouncing point. However, the
asymptotic behavior of the comoving Hubble radius makes a
difference in various bouncing models. In this regard, (1) for some
bouncing scale factors, the Hubble radius decreases monotonically
at both sides of the bounce and finally shrinks to zero size
asymptotically, which corresponds to an accelerating late-time
Universe. Therefore, in such cases, the Hubble horizon goes to
zero at large values of the cosmic time, and only for cosmic times
near the bouncing point the Hubble horizon has an infinite size.
So the primordial perturbation modes relevant for present time era
generate for cosmic times near the bouncing point, because
only at that time all the primordial modes are contained in the
horizon. As the horizon shrinks, the modes exit the horizon and
become relevant for present time observations. On the other hand,
(2) some bouncing scale factors lead to a divergent Hubble
radius asymptotically, which corresponds to a decelerating
Universe at late times. In such cases, the perturbation modes generate at
very large negative cosmic times, corresponding to the low
curvature regime of the contracting era, unlike to the previous
situations, where the perturbation modes generate near the
bouncing era. More explicitly, in the latter case, the comoving
wave number $k$ begins its propagation through spacetime at large
negative cosmic times, in the contracting phase on sub-Hubble
scales, and exits the Hubble radius during this phase, and
re-enters the Hubble radius during the low-curvature regime in
expanding phase and thus being relevant for present time
observations. Therefore, the physical picture in the two cases is
very different with regard to when the perturbation modes generate. 
However as mentioned earlier, in the present work, we
will deal with an unified scenario of bounce with late time
acceleration connected by an intermediate deceleration epoch. Thus, the
late-time Universe is characterized by an accelerating scale
factor and hence the Hubble radius should shrink to zero
asymptotically at both ``sides'' of the bounce (as the scale
factor is considered to be symmetric around the bounce). This indicates that the
perturbation modes generate near the bounce, rather than
far away from the bouncing point, because near the bouncing regime
the Hubble radius has an infinite size and all the perturbation
modes are contained inside the horizon. Hence we solve the
perturbation equations near the bounce i.e near $t = 0$, which is
the primary subject in the remaining part of this section.

\subsection{Scalar perturbation}

In principle, perturbations should always be expressed in terms of
gauge invariant quantities. In the present work, we shall work in the comoving gauge defined by the vanishing of the momentum density 
$\delta T_{0i} = 0$ (where $T_{\mu\nu}$ is the effective energy-momentum tensor and the symbol '$\delta$' denotes the corresponding perturbation). 
In this gauge, the scalar metric perturbation is expressed as,
\begin{eqnarray}
 \delta g_{ij} = a^2(t)\big[1 - 2\Re\big]\delta_{ij}
 \label{scalar per metric}
\end{eqnarray}
where $\Re(t,\vec{x})$ denotes the scalar perturbation and known as the comoving curvature perturbation which is indeed a gauge invariant quantity. 
The additional metric perturbations $\delta g_{00}$ and $\delta g_{0i}$ can be obtained in terms of $\Re$ from perturbed gravitational 
equations and as a result, the second order action of $\Re(t,\vec{x})$ is given by \cite{Hwang:2005hb,Noh:2001ia,Hwang:2002fp},
 \begin{eqnarray}
  \delta S_{\Re} = \int dt d^3\vec{x} a(t) z(t)^2\left[\dot{\Re}^2
 - \frac{1}{a^2}\left(\partial_i\Re\right)^2\right]\, ,
 \label{scalar per action}
 \end{eqnarray}
 with $z(t)$ has the following expression,
 \begin{eqnarray}
  z(t) = \frac{a(t)}{\kappa\bigg(H(t) + \frac{1}{2F'(R)}\frac{dF'(R)}{dt}\bigg)} \sqrt{\frac{3}{2F'(R)}\bigg(\frac{dF'(R)}{dt}\bigg)^2}
  \label{scalar per z}
 \end{eqnarray}
 Eq.(\ref{scalar per action}) clearly indicates that the sound speed of the scalar perturbation ($c_s^2$) is unity, which in turn
 guarantees the absence of superluminal modes or equivalently one may argue that the model is free from gradient instabilities. The unit sound
 speed for scalar perturbation is, in fact, a generic nature of $F(R)$ theory and also of a scalar-tensor theory. This equivalence,
 in respect of sound speed, between $F(R)$ and scalar tensor theory is however expected as both the theories can be mapped to one another (in the
 action level) by a conformal transformation of the metric. Coming back to the action (\ref{scalar per action}), the scalar perturbation
 has positive kinetic terms if $z^2(t) > 0$ holds, which is equivalent to the condition $F'(R) > 0$ as evident from
 Eq.(\ref{scalar per z}). We will show that the $F(R)$ obtained in the earlier sections indeed satisfies $F'(R) > 0$ and
 ensures the stability of the scalar perturbation. The action $\delta S_{\Re}$ leads to the equation for the perturbed variable
$\Re(\vec{x},t)$ as,
 \begin{eqnarray}
  \frac{1}{a(t)z^2(t)}\frac{d}{dt}\bigg[a(t)z^2(t)\dot{\Re}\bigg] - \frac{1}{a^2}\partial_{i}\partial^{i}\Re = 0
  \label{scalar per eom1}
 \end{eqnarray}
In terms of the Fourier transformed scalar perturbation variable
$\Re_k(t) = \int d\vec{x}
e^{-i\vec{k}.\vec{x}}\Re(\vec{x},t)$, the above equation can be
written as,
 \begin{eqnarray}
  \frac{1}{a(t)z^2(t)}\frac{d}{dt}\bigg[a(t)z^2(t)\dot{\Re}_k\bigg] + \frac{k^2}{a^2}\Re_k(t) = 0
  \label{scalar per eom2}
 \end{eqnarray}
where $k$ is the wave number for the $k$-th perturbation mode. As
mentioned earlier, we solve the above equation for cosmic times
near the bouncing point as the perturbation modes themselves are
generated close to the bounce and thereby we can use the form of
$F(R)$ reconstructed in Eq.(\ref{rec_bounce sol2}). With this
expression of near-bounce-$F(R)$, we will determine $z(t)$ (an
essential ingredient of perturbation equation (\ref{scalar per
eom2})) and for this purpose, we first need to evaluate $F'(R)$
and $F''(R)$ which are given by,
\begin{eqnarray}
 F'(R) = \frac{12\alpha D}{\sqrt{e}} + \frac{D}{24}\bigg[-\frac{6\big(R - 12\alpha\big)}{\sqrt{e}} +
 \frac{\sqrt{6\pi}e^{-\frac{R}{24\alpha}}\big(R - 48\alpha\big)\sqrt{R - 12\alpha}~
 Erfi\big[\frac{\sqrt{R - 12\alpha}}{2\sqrt{6\alpha}}\big]}{2\sqrt{\alpha}}\bigg]
 \label{scalar per f1_new}
\end{eqnarray}
and
\begin{eqnarray}
 F''(R) = \frac{D}{576\alpha^{3/2}}\bigg[\frac{12\sqrt{\alpha}\big(R - 72\alpha\big)}{\sqrt{e}} -
 \frac{\sqrt{6\pi}e^{-\frac{R}{24\alpha}}\big(R^2 - 96\alpha R + 1440\alpha^2\big)~
 Erfi\big[\frac{\sqrt{R - 12\alpha}}{2\sqrt{6\alpha}}\big]}{\sqrt{R - 12\alpha}}\bigg]
 \label{scalar per f2_new}
\end{eqnarray}
respectively. Recall, the Ricci scalar scalar at the bounce
becomes $R(0) = 12\alpha$ and thus the function
$Erfi\bigg[\frac{\sqrt{R - 12\alpha}}{2\sqrt{6\alpha}}\bigg]$ can
be expressed as Taylor series expansion in the powers of $\sqrt{R
- 12\alpha}$ near the bounce. However the terms with $\mathcal{O}((R -
12\alpha)^{5/2})$ in the Taylor expansion can be neglected as such
terms will eventually lead to a higher power of $t$ in comparison
to $t^2$ and all the near-bounce quantities have been or will be
evaluated up to $\mathcal{O}(t^2)$, as we also did for the near-bounce scale
factor in Sec.[\ref{sec_rec_bounce}]. Using the Taylor expansion
of $Erfi\bigg[\frac{\sqrt{R - 12\alpha}}{2\sqrt{6\alpha}}\bigg]$
along with the expression of $R(t)$ (see Eq.(\ref{rec_bounce
ricci})), we evaluate the $F'(R)$ and $F''(R)$ up to $\mathcal{O}(t^2)$ as
follows,
\begin{eqnarray}
 F'(R(t)) = \frac{12\alpha D}{\sqrt{e}} - \frac{12\alpha D}{\sqrt{e}}~\alpha t^2
 \label{scalar per f1}
\end{eqnarray}
and
\begin{eqnarray}
 F''(R(t)) = -\frac{1}{\alpha}\bigg[\frac{\alpha D}{\sqrt{e}} - \frac{4\alpha D}{3\sqrt{e}}~\alpha t^2\bigg]
 \label{scalar per f2}
\end{eqnarray}
respectively. Eq.(\ref{scalar per f1}) indicates that $F'(R)$ is positive near the bounce
 and hence ensures the stability of the scalar perturbation. On other hand, $F''(R)$ becomes negative near the bounce, which is clear from
 Eq.(\ref{scalar per f2}). Actually the condition $F''(R) < 0$ is connected with the violation of null energy condition in the $F(R)$ gravity
 theory and thus such condition should hold in order for a bounce to be generated, and the demonstration goes as follows: recall, as mentioned in
 Sec.[\ref{sec_F(R)}] that the $F(R)$ gravity contributes an effective energy-momentum tensor where the effective energy density ($\rho_{eff}$) and
the pressure ($p_{eff}$) are given in Eqs.(\ref{ed}) and
(\ref{pressure}) respectively. Using these expressions of
$\rho_{eff}$ and $p_{eff}$ along with the Hubble factor $H(t) =
2\alpha t$, it is easy to show that at the bounce $\rho_{eff} +
p_{eff}$ becomes $= 2\dot{H}\big(F'(R(t)) - 1\big) +
24\dot{H}^2F''(R(t))$. The $F'(R)$ and $F''(R)$, as we obtained
near $t = 0$ in the present context, leads to $\rho_{eff} +
p_{eff} < 0$ i.e the violation of energy condition which in turn
ensures a bouncing phenomena at $t = 0$. Plugging the expressions
of $F'(R)$, $F''(R)$ into Eq.(\ref{scalar per z}), we get,
 \begin{eqnarray}
  a(t)z^2(t) = \frac{72\alpha D}{\kappa^2\sqrt{e}}\bigg[1 + \frac{2}{3}\alpha t^2\bigg] = U_1 + V_1~\alpha t^2
  \label{scalar per z2}
 \end{eqnarray}
where $U_1 = \frac{72\alpha D}{\kappa^2\sqrt{e}}$ and
 $V_1 = \frac{48\alpha D}{\kappa^2\sqrt{e}}$. With the above expression of $a(t)z^2(t)$,
the scalar perturbed equation (\ref{scalar per eom2}) takes the
following form,
\begin{eqnarray}
 U_1\ddot{\Re}_k + 2\alpha V_1\dot{\Re}_k~t + k^2U_1\Re_k(t) = 0
 \label{scalar per eom3}
\end{eqnarray}
at leading order in $t$. Solving Eq.(\ref{scalar per eom3}) for
$\Re_k(t)$, we get,
\begin{eqnarray}
 \Re_k(t) = b_1(k)~e^{-\frac{V_1}{U_1}\alpha t^2}~H\bigg[-1 + \frac{k^2}{2\alpha}\frac{U_1}{V_1}, \sqrt{\frac{\alpha V_1}{U_1}}~t\bigg]
 \label{scalar per sol1}
\end{eqnarray}
with $H[n,x]$ is the nth order Hermite polynomial. $b_1(k)$ is the integration constant which can be determined by the initial state of the 
canonical Mukhanov-Sasaki variable $v_k(t)$ defined by $v_k(t) = z(t)\Re_k(t)$, which is the adiabatic vacuum state. Thus near the bounce 
(i.e near $t \simeq 0$), the field $v_k(t)$ satisfies
\begin{eqnarray}
 \lim_{t\rightarrow 0}v_k = \frac{1}{\sqrt{2k}}e^{-ik\tau}
 \label{BD_1}
\end{eqnarray}
where $\tau$ is the conformal time defined by $d\tau = \frac{dt}{a(t)}$. However near $t \simeq 0$, the conformal and cosmic time become same as,
\begin{eqnarray}
 \tau = \int \frac{dt}{a(t)} = \int \frac{dt}{(1 + \alpha t^2)} \simeq t~~.
 \nonumber
\end{eqnarray}
Furthermore for cosmic times that satisfy $t \simeq 0$, the Hubble horizon is infinitely large and thus the primordial modes are well inside the Hubble 
horizon i.e the perturbation modes satisfy $k \gg aH$. As a result, the equation for the field $v_k(t)$ becomes 
\begin{eqnarray}
 \frac{d^2v_k}{d\tau^2} + k^2v_k = 0~~.
 \label{BD_2}
\end{eqnarray}
This is the equation of a simple harmonic oscillator with time-independent frequency. For this case, the solution for the minimum energy state is given by 
$v_k = \frac{1}{\sqrt{2k}}e^{-ik\tau}$. Hence we impose the initial condition as shown in Eq.(\ref{BD_1}). This justifies the choice 
of the adiabatic vacuum state in the present context. On other hand, the
function $H\bigg[-1 + \frac{k^2U_1}{2\alpha V_1} ,
\sqrt{\frac{\alpha V_1}{U_1}}t\bigg]$ has a limiting value for $t
\rightarrow 0$ and is given by,
\begin{eqnarray}
  H\bigg[-1 + \frac{k^2U_1}{2\alpha V_1} , \sqrt{\frac{\alpha V_1}{U_1}}~t\bigg]\bigg|_{t \rightarrow 0} =
  \frac{2^{\frac{k^2U_1}{2\alpha V_1}}\sqrt{\pi}}{2\Gamma\big(1 - \frac{k^2U_1}{4\alpha V_1}\big)}\nonumber
\end{eqnarray}
with $\Gamma(z)$ symbolizes the ``Gamma function``. Therefore, by considering the adiabatic vacuum as the initial
condition of $\Re_k(t)$, the integration constant turns out to be,
\begin{eqnarray}
 b_1(k) = \frac{1}{z(t \rightarrow 0)}\bigg[\frac{2\Gamma\big(1 - \frac{k^2U_1}{4\alpha V_1}\big)}{\sqrt{2\pi k}~2^{\frac{k^2U_1}{2\alpha V_1}}}\bigg]
 = \frac{1}{\sqrt{U_1}}\bigg[\frac{2\Gamma\big(1 - \frac{k^2U_1}{4\alpha V_1}\big)}{\sqrt{2\pi k}~2^{\frac{k^2U_1}{2\alpha V_1}}}\bigg]
 \label{scalar per bc}
\end{eqnarray}
where in the second equality, we used $z(t\rightarrow 0) =
\frac{1}{\sqrt{U_1}}$ from Eq.(\ref{scalar per z2}). Thus as a
whole, the solution for the scalar perturbation variable becomes,
\begin{eqnarray}
 \Re_k(t)&=&\bigg(\frac{2\Gamma\big(1 - \frac{k^2U_1}{4\alpha V_1}\big)}{\sqrt{2\pi k}~2^{\frac{k^2U_1}{2\alpha V_1}}\sqrt{U_1}}\bigg)~
 e^{-\frac{V_1}{U_1}\alpha t^2}~H\bigg[-1 + \frac{k^2U_1}{2\alpha V_1}, \sqrt{\frac{\alpha V_1}{U_1}}~t\bigg]\nonumber\\
 &=&\bigg(\frac{2\kappa~\Gamma\big(1 - \frac{3k^2}{8\alpha}\big)}{\sqrt{2\pi k}~2^{\frac{3k^2}{4\alpha}}\sqrt{12\alpha D}}\bigg)~
 e^{\big[\frac{1}{4} - \frac{2}{3}\alpha t^2\big]}~H\bigg[-1 + \frac{3k^2}{4\alpha}, \sqrt{\frac{2\alpha}{3}}~t\bigg]
 \label{scalar per sol2}
\end{eqnarray}
where we have used $U_1 = \frac{72\alpha D}{\kappa^2\sqrt{e}}$ and
$V_1 = \frac{48\alpha D}{\kappa^2\sqrt{e}}$ in the second
equality. Consequently, the above solution of $\Re_k(t)$
immediately leads to the scalar power spectrum for $k$-th mode as
follows,
\begin{eqnarray}
 P_{\Re}(k,t)&=&\frac{k^3}{2\pi^2}~\bigg|\Re_k(t)\bigg|^2\nonumber\\
 &=&\frac{k^2}{12\alpha D\pi^3}~\frac{\bigg(\kappa~\Gamma\big(1 - \frac{3k^2}{8\alpha}\big)\bigg)^2}
 {~2^{\frac{3k^2}{2\alpha}}~} e^{\big[\frac{1}{2} - \frac{4}{3}\alpha t^2\big]}~
 \bigg\{H\bigg[-1 + \frac{3k^2}{4\alpha}, \sqrt{\frac{2\alpha}{3}}~t\bigg]\bigg\}^2
 \label{scalar power spectrum}
\end{eqnarray}
Using the horizon crossing relation $k = aH$, the scalar power spectrum can be expressed in terms of the quantities at horizon crossing as,
\begin{eqnarray}
 P_{\Re}(k,t)\bigg|_{h.c} = \frac{\alpha t_h^2}{3 D\pi^3}~\frac{\bigg(\kappa~\Gamma\big(1 - \frac{3\alpha t_h^2}{2}\big)\bigg)^2}
 {~2^{6\alpha t_h^2}~} e^{\big[\frac{1}{2} - \frac{4}{3}\alpha t_h^2\big]}~
 \bigg\{H\bigg[-1 + 3\alpha t_h^2, \sqrt{\frac{2\alpha}{3}}~t_h\bigg]\bigg\}^2
\label{scalar power spectrum_HC}
\end{eqnarray}
where $t_h$ is the horizon crossing time. With Eq.~(\ref{scalar power spectrum}), we can determine the
spectral index of the primordial curvature perturbations. However
before proceeding to calculate the spectral index, we will perform
first the tensor perturbation, which is necessary for evaluating
the tensor-to-scalar ratio.

\subsection{Tensor perturbation}

Let us now focus on the tensor perturbations, and the tensor
perturbation on the FRW metric background is defined as follows,
\begin{eqnarray}
 ds^2 = -dt^2 + a(t)^2\left(\delta_{ij} + h_{ij}\right)dx^idx^j\, ,
 \label{ten per metric}
\end{eqnarray}
where $h_{ij}(t,\vec{x})$ is the tensor perturbation. The variable $h_{ij}(t,\vec{x})$
is itself a gauge invariant quantity, and the tensor perturbed action up to quadratic order is given by,
\begin{eqnarray}
 \delta S_{h} = \int dt d^3\vec{x} a(t) z_T(t)^2\left[\dot{h}_{ij}\dot{h}^{ij}
 - \frac{1}{a^2}\left(\partial_lh_{ij}\right)^2\right]\, ,
 \label{ten per action}
\end{eqnarray}
where $z_T(t)$ has the following form \cite{Hwang:2005hb},
\begin{eqnarray}
 z_T(t) = \frac{a(t)}{\kappa}\sqrt{F'(R)}\, ,
 \label{ten per z}
\end{eqnarray}
Therefore the speed of the tensor perturbation is $c_T^2 = 1$ i.e
the gravitational waves propagate with the speed of light which is
unity in the natural units.

Coming back to Eq.(\ref{ten per z}), the stability of the tensor
perturbation will be guaranteed by the condition $F'(R) > 0$ - the
same condition by which the scalar perturbation is ensured to be
stable. Recall, in the previous section, we showed the positivity
of $F'(R)$ in Eq. (\ref{scalar per f1}) and thus the tensor
perturbation is indeed stable in the present context. The action
(\ref{ten per action}) leads to the equation of the tensor
perturbed variable as,
\begin{eqnarray}
  \frac{1}{a(t)z_T^2(t)}\frac{d}{dt}\bigg[a(t)z_T^2(t)\dot{h}_{ij}\bigg] - \frac{1}{a^2}\partial_{l}\partial^{l}h_{ij} = 0
  \label{ten per eom1}
 \end{eqnarray}
The Fourier transformed tensor perturbation variable is defined as $h_{ij}(t,\vec{x}) = \int d\vec{k}~\sum_{\gamma}\epsilon_{ij}^{(\gamma)}~
h_{(\gamma)}(\vec{k},t) e^{i\vec{k}.\vec{x}}$, where $\gamma = '+'$ and $\gamma = '\times'$ represent two polarization modes. Moreover
$\epsilon_{ij}^{(\gamma)}$ are the polarization tensors satisfying $\epsilon_{ii}^{(\gamma)} = k^{i}\epsilon_{ij}^{(\gamma)} = 0$. In
terms of the Fourier transformed tensor variable $h_k(t)$, Eq.(\ref{ten per eom1}) can be expressed as,
 \begin{eqnarray}
  \frac{1}{a(t)z_T^2(t)}\frac{d}{dt}\bigg[a(t)z_T^2(t)\dot{h}_k\bigg] + \frac{k^2}{a^2}h_k(t) = 0
  \label{ten per eom2}
 \end{eqnarray}
 The two polarization modes obey the same Eq.(\ref{ten per eom2}) and thus we omit the polarization index. Moreover, both the polarization modes
 even follow the same initial condition and that is why they have the same solution. So in the following, what we will do in the
 context of tensor perturbation is applicable for both the polarization modes. Finally, in the expression of the tensor power spectrum, we will
 introduce a multiplicative factor $2$ due to the contribution from both the polarization modes. Using the expression of $F'(R)$ from
 Eq.(\ref{scalar per f1}), we determine $a(t)z_T^2(t)$ as,
 \begin{eqnarray}
  a(t)z_T^2(t) = \frac{12\alpha D}{\kappa^2\sqrt{e}}\bigg[1 + 2\alpha t^2\bigg] = U_2 + V_2~\alpha t^2
  \label{ten per z2}
 \end{eqnarray}
with $U_2 = \frac{12\alpha D}{\kappa^2\sqrt{e}}$ and $V_2 =
\frac{24\alpha D}{\kappa^2\sqrt{e}}$. Plugging back this
expression of $a(t)z_T^2(t)$ into Eq.(\ref{ten per eom2}) and
after some algebra, we get the following equation,
\begin{eqnarray}
 U_2\ddot{h}_k + 2\alpha V_2\dot{h}_k~t + k^2U_2h_k(t) = 0
 \label{ten per eom3}
\end{eqnarray}
at leading order in $t$, which is a viable consideration because
the perturbation modes generate near the bouncing phase in the
present scenario. Solving Eq.(\ref{ten per eom3}) for $h_k(t)$ ,
we get,
\begin{eqnarray}
 h_k(t) = b_2(k)~e^{-\frac{V_2}{U_2}\alpha t^2}~H\bigg[-1 + \frac{k^2}{2\alpha}\frac{U_2}{V_2}, \sqrt{\frac{V_2}{U_2}\alpha}~t\bigg]
 \label{ten per sol1}
\end{eqnarray}
where $b_2(k)$ is an integration constant and can be determined from an initial condition. As an initial condition, we consider that
the tensor perturbation field starts from the adiabatic vacuum, more precisely the initial configuration is given by,
$\lim_{t\rightarrow 0}\big[z_T(t)h_k(t)\big] = \frac{1}{\sqrt{2k}}$. This immediately leads to the expression of $b_2(k)$ as,
\begin{eqnarray}
 b_2(k) = \frac{1}{z_T(t \rightarrow 0)}\bigg[\frac{2\Gamma\big(1 - \frac{k^2U_2}{4\alpha V_2}\big)}{\sqrt{2\pi k}~2^{\frac{k^2U_2}{2\alpha V_2}}}\bigg]
 = \frac{1}{\sqrt{U_2}}\bigg[\frac{2\Gamma\big(1 - \frac{k^2U_2}{4\alpha V_2}\big)}{\sqrt{2\pi k}~2^{\frac{k^2U_2}{2\alpha V_2}}}\bigg]~~~.
 \label{ten per bc}
\end{eqnarray}
In the second equality of the above equation, we use $z_T(t\rightarrow 0) = 1/\sqrt{U_2}$ from Eq.(\ref{ten per z2}). Plugging
this expression of $b_2(k)$ into Eq.(\ref{ten per sol1}) yields the final solution of $h_k(t)$ as follows,
\begin{eqnarray}
 h_k(t)&=&\bigg(\frac{2\Gamma\big(1 - \frac{k^2U_2}{4\alpha V_2}\big)}{\sqrt{2\pi k}~2^{\frac{k^2U_2}{2\alpha V_2}}\sqrt{U_2}}\bigg)~
 e^{-\frac{V_2}{U_2}\alpha t^2}~H\bigg[-1 + \frac{k^2U_2}{2\alpha V_2}, \sqrt{\frac{V_2}{U_2}\alpha}~t\bigg]\nonumber\\
 &=&\bigg(\frac{2\kappa~\Gamma\big(1 - \frac{k^2}{8\alpha}\big)}{\sqrt{2\pi k}~2^{\frac{k^2}{4\alpha}}\sqrt{12\alpha D}}\bigg)~
 e^{\big[\frac{1}{4} - 2\alpha t^2\big]}~H\bigg[-1 + \frac{k^2}{4\alpha}, \sqrt{2\alpha}~t\bigg]
 \label{ten per sol2}
\end{eqnarray}
where we have used $U_2 = \frac{12\alpha D}{\kappa^2\sqrt{e}}$ and $V_2 = \frac{24\alpha D}{\kappa^2\sqrt{e}}$.
Eq.(\ref{ten per sol2}) represents the solution of the tensor perturbation for both the polarization modes. The solution of $h_k(t)$ immediately
leads to the tensor power spectrum as,
\begin{eqnarray}
 P_{h}(k,t)&=&\frac{k^3}{2\pi^2}~\sum_{\gamma}\bigg|h_k^{(\gamma)}(t)\bigg|^2\nonumber\\
 &=&\frac{k^2}{6\alpha D\pi^3}~\frac{\bigg(\kappa~\Gamma\big(1 - \frac{k^2}{8\alpha}\big)\bigg)^2}
 {~2^{\frac{k^2}{2\alpha}}} e^{\big[\frac{1}{2} - 4\alpha t^2\big]}~
 \bigg\{H\bigg[-1 + \frac{k^2}{4\alpha}, \sqrt{2\alpha}~t\bigg]\bigg\}^2
 \label{ten power spectrum}
\end{eqnarray}
It may be noticed that $\gamma = '+'$ and $\gamma = '\times'$ modes contribute equally to the power spectrum, as expected because their
solutions behave similarly. At the horizon crossing, the tensor power spectrum turns out to be,
\begin{eqnarray}
 P_{h}(k,t)\bigg|_{h.c} = \frac{2\alpha t_h^2}{3 D\pi^3}~\frac{\bigg(\kappa~\Gamma\big(1 - \frac{\alpha t_h^2}{2}\big)\bigg)^2}
 {~2^{2\alpha t_h^2}} e^{\big[\frac{1}{2} - 4\alpha t_h^2\big]}~
 \bigg\{H\bigg[-1 + \alpha t_h^2, \sqrt{2\alpha}~t_h\bigg]\bigg\}^2
 \label{ten power spectrum_H.C}
\end{eqnarray}

\begin{figure}[!h]
\begin{center}
 \centering
 \includegraphics[width=3.5in,height=2.0in]{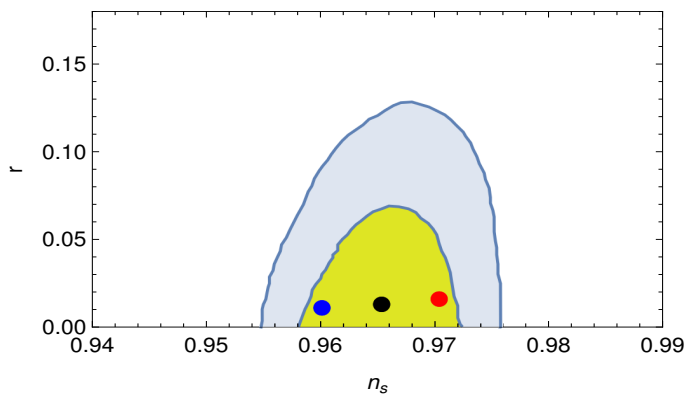}
 \caption{$1\sigma$ (yellow) and $2\sigma$ (light blue) contours for Planck 
 2018 results \cite{Akrami:2018odb}, on $n_s-r$ plane. 
 Additionally, we present the predictions of the present bounce scenario with $\frac{R_h}{\alpha} = 14$ (blue point), $\frac{R_h}{\alpha} = 16$ 
 (black point) and $\frac{R_h}{\alpha} = 20$ (red point).}
 \label{plot_observable}
\end{center}
\end{figure}

Now we can explicitly confront the model at hand with the latest
Planck observational data \cite{Akrami:2018odb}, so we shall
calculate the spectral index of the primordial curvature
perturbations $n_s$ and the tensor-to-scalar ratio $r$, which are defined as follows,
\begin{eqnarray}
 n_s - 1 = \frac{\partial\ln{P_{\Re}}}{\partial\ln{k}}\bigg|_{H.C}~~~~~~~~,~~~~~~~~~~~
 r = \frac{P_h(k,t)}{P_{\Re}(k,t)}\bigg|_{H.C}\label{spectral index1}
\end{eqnarray}
As evident from these expressions, $n_s$ and $r$ are evaluated at
the time of the first horizon exit near the bouncing point, for
positive times (symbolized by 'H.C' in the above equations), when
$k=aH$ i.e. when the mode $k$ crosses the Hubble horizon. Using
Eq.(\ref{scalar power spectrum}), we determine
$\frac{\partial\ln{P_{\Re}}}{\partial\ln{k}}$ as follows,
\begin{eqnarray}
 \frac{\partial\ln{P_{\Re}}}{\partial\ln{k}} = 2 - \frac{3k^2}{2\alpha}~\psi^{(0)}\bigg[1 - \frac{3k^2}{8\alpha}\bigg]
 - \frac{3k^2}{\alpha}(\ln{2}) + \frac{3k^2}{\alpha}\bigg\{\frac{H^{(1,0)}
 \bigg[-1 + \frac{3k^2}{4\alpha}, \sqrt{\frac{2\alpha}{3}}~t\bigg]}
 {H\bigg[-1 + \frac{3k^2}{4\alpha}, \sqrt{\frac{2\alpha}{3}}~t\bigg]}\bigg\}
\label{spectral index2}
\end{eqnarray}
where $\psi^{(0)}[z]$ is the zeroth order Polygamma function or equivalently the digamma function and
$H^{(1,0)}[z_1,z_2]$ is the derivative of
$H[z_1,z_2]$ with respect to its first argument. Moreover
the following expressions were used to calculate Eq.(\ref{spectral
index2}),
\begin{eqnarray}
 \frac{\partial}{\partial k}\bigg\{\Gamma\bigg(1 - \frac{3k^2}{8\alpha}\bigg)\bigg\}&=&
 -\frac{3k}{4\alpha}~\Gamma\bigg(1 - \frac{3k^2}{8\alpha}\bigg)~\psi^{(0)}\bigg(0, 1 - \frac{3k^2}{8\alpha}\bigg)\nonumber\\
 \frac{\partial}{\partial k}\bigg\{H\bigg[-1 + \frac{3k^2}{4\alpha}, \sqrt{\frac{2\alpha}{3}}~t\bigg]\bigg\}&=&
 \frac{3k}{2\alpha}~H^{(1,0)}\bigg[-1 + \frac{3k^2}{4\alpha}, \sqrt{\frac{2\alpha}{3}}~t\bigg]
 \label{new2}
\end{eqnarray}
Thereby Eq.(\ref{spectral index2}) immediately leads to the spectral index as,
\begin{eqnarray}
 n_s = \bigg[3 - \frac{3k^2}{2\alpha}~\psi^{(0)}\bigg(0, 1 - \frac{3k^2}{8\alpha}\bigg)
 - \frac{3k^2}{\alpha}(\ln{2}) + \frac{3k^2}{2\alpha}\bigg\{\frac{H^{(1,0)}\bigg(-1 + \frac{3k^2}{4\alpha}, \sqrt{\frac{2\alpha}{3}}~t\bigg)}
 {H\bigg(-1 + \frac{3k^2}{4\alpha}, \sqrt{\frac{2\alpha}{3}}~t\bigg)}\bigg\}\bigg]_{H.C}
 \label{spectral index3}
\end{eqnarray}
As mentioned earlier, the perturbation modes are generated and
also cross the horizon near the bounce. Thus we can safely use the
near-bounce scale factor in the horizon crossing condition to
determine $k = aH = 2\alpha t_h$ (where $t_h$ is the horizon
crossing time). Using this relation, Eq.(\ref{spectral index3})
turns out to be,
\begin{eqnarray}
 n_s = 3 - 6\alpha t_h^2~\psi^{(0)}\bigg(0, 1 - \frac{3}{2}~\alpha t_h^2\bigg)
 - 12\alpha t_h^2(\ln{2}) + 12\alpha t_h^2
 \bigg\{\frac{H^{(1,0)}\bigg(-1 + 3\alpha t_h^2, \sqrt{\frac{2\alpha}{3}}~t_h\bigg)}
 {H\bigg(-1 + 3\alpha t_h^2, \sqrt{\frac{2\alpha}{3}}~t_h\bigg)}\bigg\}
 \label{spectral index4}
\end{eqnarray}
Furthermore, the tensor-to-scalar ratio is given by,
\begin{eqnarray}
 r = \bigg|\frac{h_k(t)}{\Re_k(t)}\bigg|^2_{k=a(t_h)H(t_h)}
 \label{tensor to scalar ratio}
\end{eqnarray}
where the solutions of $h_k(t)$ and $\Re_k(t)$ are shown in
Eqs.(\ref{ten per sol2}) and (\ref{scalar per sol2}) respectively.
Eqs.(\ref{spectral index4}) and (\ref{tensor to scalar ratio})
clearly indicate that $n_s$ and $r$ depend on the dimensionless
parameter $\alpha t_h^2$ which is further connected to the Ricci
scalar at horizon crossing by $\alpha t_h^2 =
\big(\frac{R_h}{12\alpha} - 1\big)$. Thereby, we can argue that
the observable quantities $n_s$ and $r$ depend on $R_h/\alpha$.
With this information, we now directly confront the theoretical expectations of spectral index
and tensor-to-scalar ratio with the Planck 2018 constraints
\cite{Akrami:2018odb}. In Fig.~[\ref{plot_observable}] we present the estimated spectral index and 
tensor-to-scalar ratio of the present scenario for three choices of $\frac{R_h}{\alpha}$, on top of the $1\sigma$ and $2\sigma$ 
contours of the Planck 2018 results \cite{Akrami:2018odb}. As we observe, the agreement with observations is efficient, and in 
particular well inside the $1\sigma$ region. At this stage it is
worth mentioning that in an $F(R)$ gravity theory, the matter
bounce scenario, in which case the perturbations are generated far
away from the bouncing point deeply in the contracting regime, is
not consistent with the Planck results as shown in
\cite{Odintsov:2014gea}. However, here we demonstrate that a
$F(R)$ gravity model indeed leads to a viable bouncing model where
the primordial perturbations are generated near the bounce.
Thereby, we can argue that the viability of a $F(R)$ model (with
respect to the Planck constraints) is interrelated with the
generation era of the perturbations modes.
Furthermore the scalar perturbation amplitude ($A_s$) is
constrained by $\ln{\big[10^{10}A_s\big]} = 3.044 \pm 0.014$ due
to the Planck results \cite{Akrami:2018odb}. Eq.(\ref{scalar power
spectrum}) along with the consideration of the integration
constant $D = \frac{1}{\alpha}$ (recall $D$ has mass dimension
[-2]) depict that the scalar perturbation amplitude depends on the
dimensionless parameters $\frac{{R_h}}{\alpha}$ and
$\alpha\kappa^2$. We may choose $\frac{R_h}{\alpha} = 16$ (which
is within the range that makes the $n_s$ and $r$ compatible with
the Planck results) and consequently the scalar perturbation
amplitude in the present context becomes $A_s =
\frac{1}{20\pi^3}\alpha\kappa^2$. Therefore the theoretical
expectation of $A_s$ will match with the Planck observations
provided $\alpha\kappa^2$ lies within $\alpha\kappa^2 =
[1.283\times10^{-7} , 1.320\times10^{-7}]$. Thus as a whole, the
observable quantities $n_s$, $r$ and $A_s$ are simultaneously
compatible with the Planck constraints for the parameter ranges :
$14 \leq \frac{R_h}{\alpha} \leq 20$ and $\alpha\kappa^2 =
[1.283\times10^{-7} , 1.320\times10^{-7}]$ respectively. However
the viable range of $\alpha\kappa^2$ depends on the consideration
$D = 1/\alpha$, i.e. a different $D$ will eventually lead to a
different range of viability of the parameter $\alpha\kappa^2$. As
an example, for $D = \frac{1}{4\alpha}$, the scalar perturbation
amplitude becomes $A_s = \frac{1}{5\pi^3}\alpha\kappa^2$ while
$n_s$ and $r$ have the same form as given in Eqs.(\ref{spectral
index4}) and (\ref{tensor to scalar ratio}) respectively and
therefore the parameters should lie within $14 \leq \frac{R_h}{\alpha} \leq 20$ and 
$\alpha\kappa^2 = [3.208\times10^{-8} , 3.300\times10^{-8}]$ in order to make the
theoretical values of the observable quantities compatible with
the latest Planck 2018 results. Such parametric ranges make the
horizon crossing Ricci scalar of the order $R_h \sim
10^{-8}/\kappa^2 = 10^{28}\mathrm{GeV}^2$.

Thereby, the viable ranges of the early and late stage parameters
are given by $14 \lesssim \frac{R_h}{\alpha} \lesssim 20$,
$\alpha\kappa^2 = [3.208\times10^{-8} , 3.300\times10^{-8}]$,
$\beta = 3.98^{+\infty}_{-2.46}$ and $\Lambda = 1.2\times10^{-84}\mathrm{GeV}^2$ 
respectively. As a result, we can interpolate the scale
factor or the Hubble radius in the two transition eras (i.e from
the bounce to matter dominated and from the matter dominated to
dark energy epoch), leading to a smooth evolution of the Hubble
radius for a large range of cosmic time.\\

Before moving to the next section we would like to mention that in the present context, the non-singular bounce universe is time symmetric 
and the initial adiabatic vacuum condition 
is set at the bouncing point, which seems that the model is more similar to a scenario of the creation of the universe from nothing. 
The spontaneous birth of the universe from ``nothing'' has been discussed earlier in the context of quantum cosmology where the universe 
is described by a wave function satisfying the well known Wheeler-DeWitt equation \cite{Halliwell:1990uy}. With the development of 
quantum cosmology theory, it has 
been suggested that the universe can be created spontaneously from nothing, where ``nothing'' means there is 
neither matter nor space or time, and the problem of singularity can be avoided naturally \cite{Atkatz:1994hy,He:2014yia}. 
In particular, the author of \cite{Atkatz:1994hy} showed that 
a quantum universe of zero size, or, indeed, the cosmological ``nothing'' may quantum mechanically tunnel to a universe of a finite size. However 
the argument of \cite{Atkatz:1994hy} was established for a closed type universe, unlike to the present bounce scenario 
which has been presented for a flat FRW universe.

\section{Numerical solutions: An unified description from bounce-to-deceleration-to-late time acceleration}\label{sec_unification}

Given the structure of the scale factor in the early and late
stages of the Universe (as $a_b(t) = 1+\alpha t^2$ in
Eq.(\ref{rec_bounce scale}) and $a_l(t) = e^{\int H_l(t)dt}$ respectively, where
the $H_l(t)$ is given in Fig.[\ref{plot_rec_late}]), we would like
to provide a complete picture by considering the intermediate
region to be a matter dominated epoch, and moreover the transitions
from the bounce to matter dominated (MD) and from the matter
dominated to the DE era will be obtained by the method of
numerical interpolation. Due to complicated nature of the
equations governing the evolution of the scale factor in a general
context, we will determine the interpolating function using
numerical techniques and will illustrate the same. Let us briefly
point out the methods one may use in order to generate such
interpolating solutions. In the transition regions (i.e. from the
bounce-MD and from the MD-DE), one approximates the behavior of
the physical quantity of interest (e.g., the scale factor $a(t)$
or the Hubble radius $r_h = \frac{1}{aH}$) by a polynomial
function of time, with degree of the polynomial kept arbitrary.
Then, in the early bounce epoch one uses the given behavior of the
desired physical quantity (for example for the scale factor, the
behavior near the bounce is $a_b(t) = 1+\alpha t^2$) to generate
numerical estimates of the respective quantity at various time
instants till the description is reliable. Similar numerical
estimations are being made at the intermediate matter dominated
and at the late stage as well. With these sets of data and the
polynomial function one can use any standard interpolation
software package to end up getting the desired plots. The
structure of the plot, of course, depends on the degree of the
polynomial and desired accuracy level. All the plots in this paper
are for an accuracy level of $\mathcal{O}(10^{-8})$. Since our main aim is
to merge certain cosmological epochs of the Universe, a good
physical quantity to start with is the Hubble radius ($r_h(t)$)
rather than the scale factor, because the accelerating or
decelerating stage of the Universe is easily realized by the
decreasing or increasing behavior of the Hubble radius (with respect to cosmic time)
respectively. Following Eq.(\ref{scale_ansatz}), we may construct
the Hubble radius as,
\begin{eqnarray}
 r_h(t)&=&\big[\dot{a}_b(t)\big]^{-1} = \frac{1}{2\alpha t}~~,~~~~~~~~~~~~~~~~~~~~~~~~~~~~~~~~near~the~bounce,~for~~~~~~~t\simeq 0^{+}\nonumber\\
 r_h(t)&=&\big[\dot{a}_I(t)\big]^{-1} = \frac{3t}{2}\bigg(\frac{T}{t}\bigg)^{2/3}~~,~~~~~~~in~the~intermediate~regime,~for~~~~~~~0^{+} \lesssim t \lesssim 5\times10^{17}~sec.\nonumber\\
 r_h(t)&=&\big[\dot{a}_l(t)\big]^{-1}~~~(see~Fig.[\ref{plot_rec_late}])~~,~~~~~~~~~~~~~~~~~~~~~during~late~time,~for~~~~~~~~t \gtrsim 6\times10^{17}~sec.
 \label{hubble radius_ansatz}
\end{eqnarray}
and then with the procedure described above, we numerically
interpolate the Hubble radius in the two transitions i.e. from the
bounce to matter dominated and from the matter dominated to DE era
respectively. Moreover the Hubble radius in the contracting regime
can also be obtained by assuming a symmetric behavior of
$\big|r_h(t)\big|$ around the bounce. However, the details of the
interpolation of the curve connecting the early bounce to the late
accelerating stage is an artifact of the procedure followed and
admits possible variations depending on the process of
interpolation by numerical techniques. However 
such indeterminacy in determining the interpolating
function would not affect our main conclusions in the present
context. Thus, having explained the details of the interpolating
procedure, we now turn to the corresponding implications and
present the variations of all the relevant parameters with time.

\begin{figure}[!h]
\begin{center}
 \centering
 \includegraphics[width=18pc]{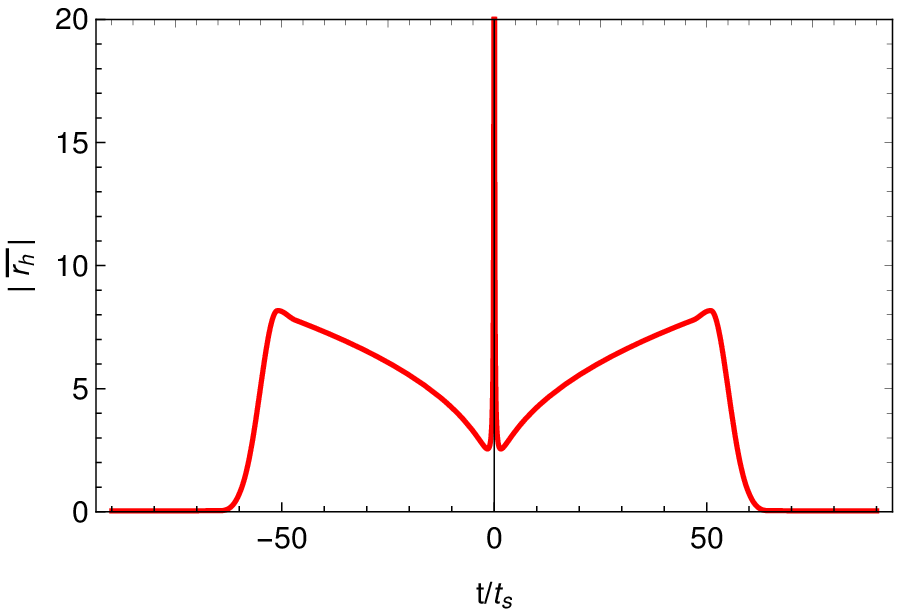}
 \includegraphics[width=18pc]{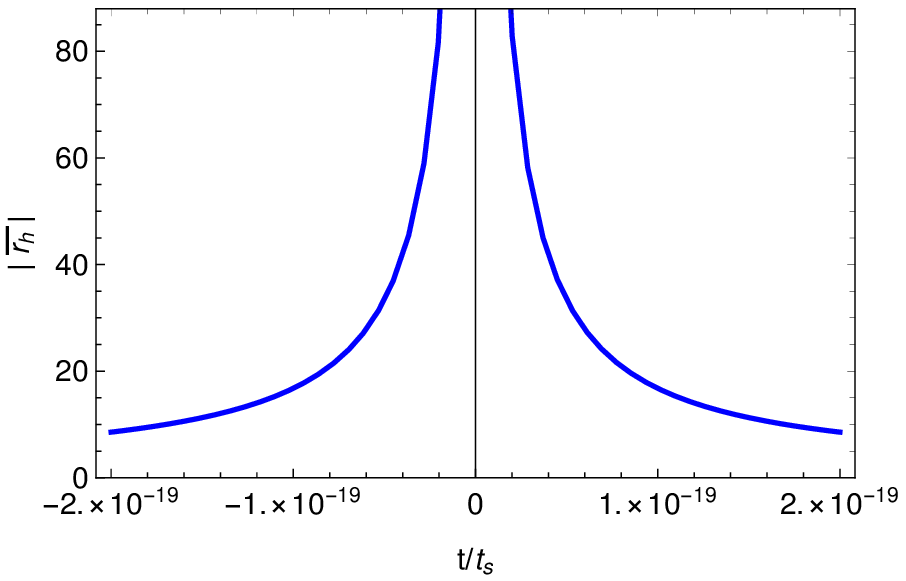}
 \caption{$Left~plot$ : The modulus of the rescaled Hubble radius $\big|\bar{r}_h\big|$ (along y-axis, in the unit of second) is being
 plotted against $\frac{t}{t_s}$ (along x-axis) for a wide range of cosmic time, in particular for $-9\times10^{17} \leq t \leq 9\times10^{17}$ sec. Such
 construction of the Hubble radius depends on Eq.(\ref{hubble radius_ansatz}) and the numerical interpolation in the transition eras i.e from
 the bounce-MD and the MD-dark energy epoch. During numerical interpolation, we take
 $\alpha\kappa^2 = 3.25\times10^{-8}$, $\beta = 4$ and $T = 1$ sec.
 The curve explicitly shows a smooth unification of the Universe evolution from the bounce-to-deceleration-to-late time acceleration.
 $Right~plot$ : A zoomed-in version of the left plot near $t = 0$.}
 \label{plot_Hubble_horizon}
\end{center}
\end{figure}

\begin{figure}[!h]
\begin{center}
 \centering
 \includegraphics[width=3.5in,height=2.0in]{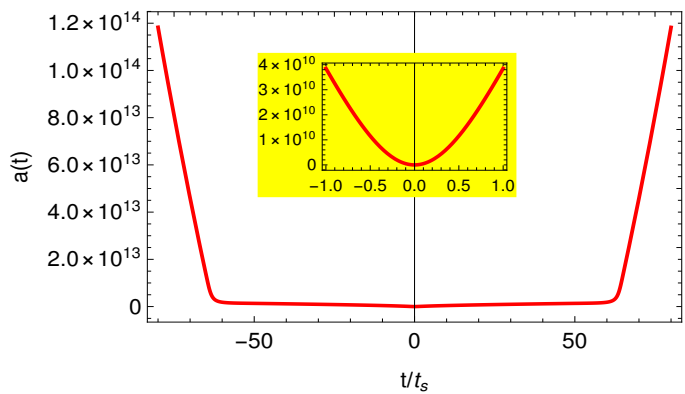}
 \caption{$a(t)$ (along y-axis) vs. $\frac{t}{t_s}$ (along x-axis).
 The scale factor gets a minimum at $t = 0$ and thus indicates a non-singular bounce at that point of time.
 To get a better view of what is happening near the bounce, we give a zoomed-in version by the inset-graph depicting the behavior of the scale factor near
 $t = 0$.}
 \label{plot_scale}
\end{center}
\end{figure}

As mentioned earlier, we start with the Hubble radius ($r_h =
\frac{1}{\dot{a}}$), in particular, taking the reduced Planck mass to be
$M_\mathrm{Pl} = 10^{18}\mathrm{GeV}$ and using the aforementioned construction of
the Hubble radius, we obtain $r_h(t)$ as a function of time in the
expanding phase of the Universe. Further the Hubble radius in the contracting
regime is obtained by considering a symmetric
character of $\big|r_h(t)\big|$ in the both sides of the bounce.
As a result, we get the Hubble radius for $-9\times10^{17} \leq t
\leq 9\times10^{17}$ sec., which is presented in the left part of
Fig.[\ref{plot_Hubble_horizon}], while the right part is a
zoomed-in version of the left one near $t = 0$. The x-axis of the
Fig.[\ref{plot_Hubble_horizon}] corresponds to a ``rescaled'' time
coordinate obtained as $\frac{t}{t_s}$ (with $t_s = 10^{16}$ sec.)
which is dimensionless, while the $y$ axis represents the rescaled
Hubble radius $\bar{r}_h = 10^{-5}r_h$ which is in the unit of ``sec''. 
Fig.[\ref{plot_Hubble_horizon}] clearly demonstrates that
the Hubble radius depicts a bounce at $t = 0$ (as expected,
because it is constructed from $r_h^{(b)}(t) = \big(a_bH_b\big)^{-1}$
near $t = 0$), then it increases monotonically with time up to
$\frac{t}{t_s} \simeq 50$ (or $t = 5\times10^{17}$ sec.) leading
to a decelerating phase of the Universe and finally after $t
\gtrsim 5\times10^{17}$ sec., the Hubble radius starts to
decrease, which in turn indicates the dark energy epoch of the
Universe. This may provide a merging of a non-singular bounce to matter dominated epoch, followed 
by a late time dark energy era. It may be noticed
that the Hubble radius goes to zero asymptotically at both
``sides'' of the bounce, which in turn confirms the fact that the
relevant primordial perturbation modes generate near the
bouncing point as we have considered in Sec.[\ref{sec_estimation}]
during the calculation of scalar and tensor perturbations. The
generation era of perturbation modes in the present context makes
the model different than the usual matter bounce scenario where
the perturbation modes generate in the distant past far away
from the bounce.
Having this Hubble radius, we are now going to solve the scale
factor and Hubble parameter, however numerically and then by using
the numerical solution of $H(t)$ we reconstruct the form of $F(R)$
from the gravitational Eq.(\ref{basic4}). The scale factor can be
obtained from $\dot{a}(t) = \frac{1}{r_h(t)}$ which is a first order
differential equation and thus the solution of the same requires one
boundary condition. We choose $a(0) = 1$, because we want the
scale factor to behave like $a_b(t)$ near $t = 0$ and recall
$a_b(0) = 1$. With such initial condition along with the $r_h(t)$
given in Fig.[\ref{plot_Hubble_horizon}], we solve $a(t)$
numerically, which is presented in Fig.[\ref{plot_scale}] where
once again,
the x axis is rescaled by $t_r = t/t_s$.\\
It is evident that the scale factor is smooth everywhere and the
graph in the inset shows that the scale factor acquires a minimum
at $t = 0$. Thus we can argue that the Universe transits from a contracting phase to an
expanding one through a non-singular bounce at $t = 0$  and thus the Universe evolution becomes free of the
initial singularity. The above numerical solutions of the scale
factor can be immediately differentiated providing the Hubble
parameter as a function of time, shown in Fig.[\ref{plot_Hubble
parameter}].
\begin{figure}[!h]
\begin{center}
 \centering
 \includegraphics[width=18pc]{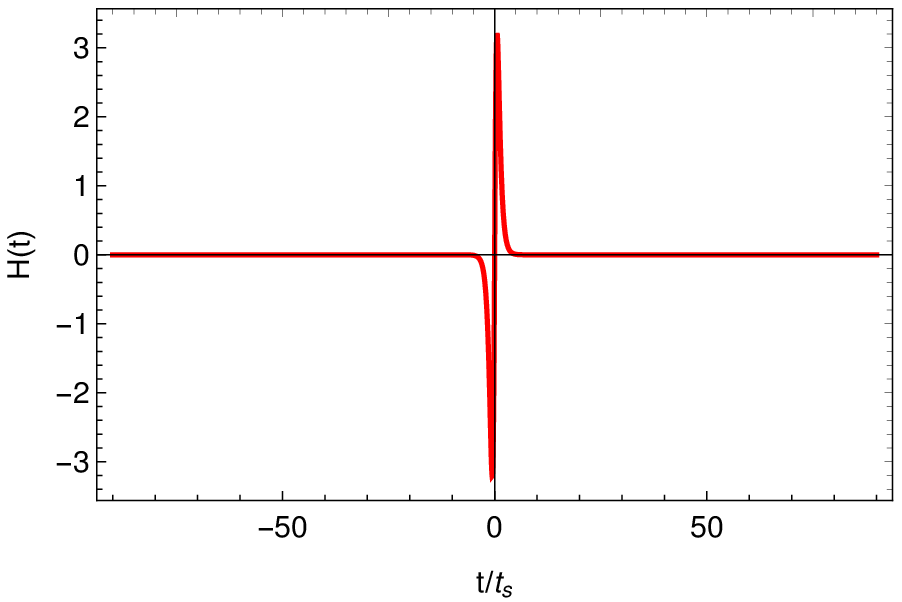}
 \includegraphics[width=18pc]{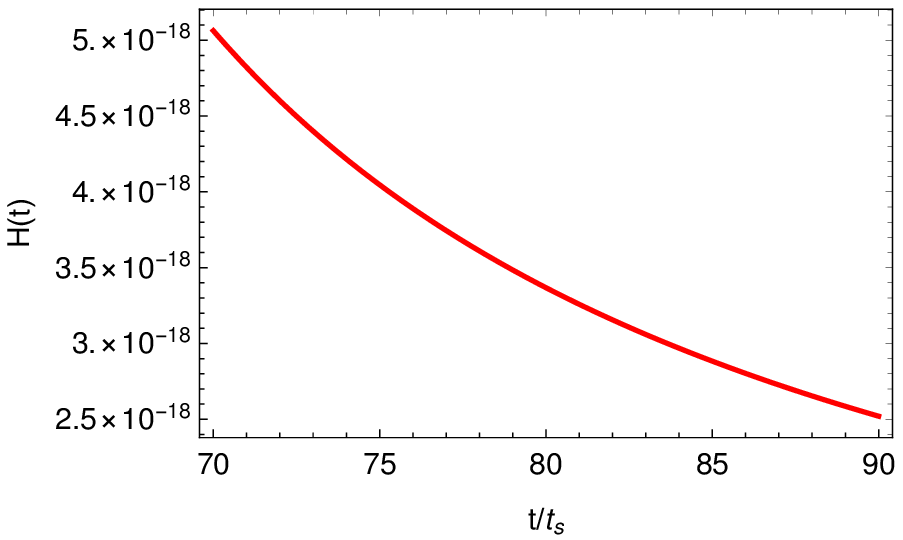}
 \caption{$Left~plot$ : $H(t)$ (along y-axis, in the unit of sec$^{-1}$) vs. $\frac{t}{t_s}$ (along x-axis)
 for $-9\times10^{17} \leq t \leq 9\times10^{17}$ sec., which is
 obtained from the solution of the scale factor shown in Fig.[\ref{plot_scale}]. As evident, $H(t)$ becomes zero at $t = 0$ and $\dot{H}(0)>0$,
 confirming a bouncing Universe. Moreover the Hubble parameter acquires a maximum near the bounce.
 Finally in the late time epoch, the Hubble parameter monotonically decreases with time.
 $Right~plot$ : A zoomed-in version of $H(t)$ vs. $\frac{t}{t_s}$ for $7\times10^{17} \leq t \leq 9\times10^{17}$ sec.}
 \label{plot_Hubble parameter}
\end{center}
\end{figure}
The Hubble parameter becomes zero and shows an increasing behavior
with time (i.e $\dot{H} > 0$) at $t = 0$, however this is expected
as the scale factor itself acquires a minimum at $t = 0$ i.e. a
bounce. Moreover, the Hubble parameter acquires a maximum very
near to bounce and then decreases rapidly, and continued to
decrease until the late-time era. A zoomed-in version of the late
time behavior of $H(t)$ is shown in the right plot of Fig.
[\ref{plot_Hubble parameter}], which clearly demonstrates that at
the late epoch i.e. for $t \gtrsim 7\times10^{17}$ sec., the
Hubble parameter becomes of the order $\sim 10^{-18}$ sec$^{-1}$.
The occurrence of the maximum of $H(t)$ near the bounce in the present
context is in agreement with \cite{Odintsov:2016tar} where the
authors explained the merging of bounce with late-time
acceleration from a different viewpoint, namely from the holonomy
generalizations of $deformed$ matter bounce scenario. In this regard, it
may be mentioned that a deformed matter bounce scenario is
represented by an asymmetric scale factor where the Hubble radius
diverges at $t \rightarrow -\infty$ and asymptotically goes to
zero at $t \rightarrow +\infty$. Thus the primordial perturbation
modes in a deformed matter bounce model generate deeply in
the contracting regime far away from the bounce, unlike to our
present case where the Hubble radius goes to zero asymptotically
at both ``sides'' of the bounce (see
Fig.[\ref{plot_Hubble_horizon}]) leading to the generation of the
perturbation modes near the bouncing regime.

\begin{figure}[!h]
\begin{center}
 \centering
 \includegraphics[width=18pc]{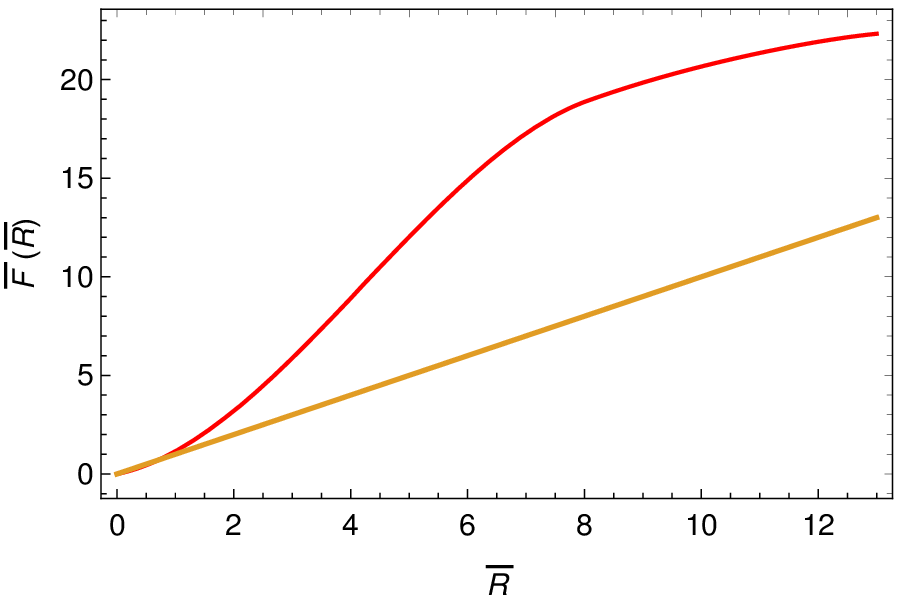}
 \includegraphics[width=18pc]{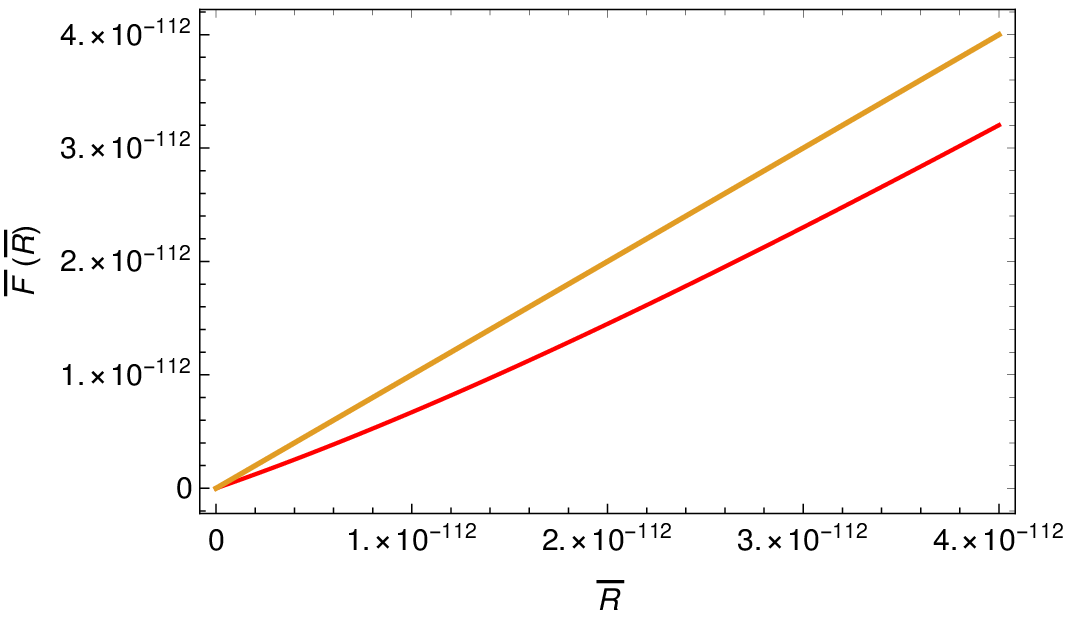}
 \caption{$Left~plot$ : $\bar{F}(\bar{R})$ (along y axis) vs. $\bar{R}$ (along x axis). The red curve depicts the numerical solution of the $F(R)$ and
 the yellow curve represents the Einstein gravity. The figure reveals that the $F(R)$ matches with the Einstein gravity in the low curvature regime,
 while the $F(R)$ seems to deviate from the usual General Relativity when the scalar curvature takes larger and larger values
 $Right~plot$ : A zoomed-in version of the left plot near $R \rightarrow 0$ to get a better view of the behaviour of the $F(R)$ during the present epoch.}
 \label{plot_fR_solution}
\end{center}
\end{figure}

With the Hubble parameter at hand, our next task is to solve the
gravitational equation (\ref{basic4}) to reconstruct the form of
$F(R)$ which can realize such evolution of the Hubble parameter.
For this purpose, we need two boundary conditions as
Eq.(\ref{basic4}) is a second order differential equation with
respect to the Ricci scalar. As mentioned earlier, we use the
forms of $F_b(R)$ and $F_l(R)$ obtained in
Sec.[\ref{sec_reconstruction}] for such boundary conditions and
thus the boundary conditions are given by: $F(R=0) = F_l(0)$ and
$F(R_b=12\alpha) = F_b(12\alpha)$ respectively (recall, $R_b$ is
the Ricci scalar at the bounce). As a result, we solve Eq.
(\ref{basic4}) numerically and the form of $F(R)$ we get is
depicted in the left part of Fig.[\ref{plot_fR_solution}].
Actually the $F(R)$ is demonstrated by the red curve, while the
yellow one represents the Einstein gravity.
Before going into the discussion regarding the behavior of the
$F(R)$ gravity, let us briefly comment about the range of the
``rescaled Ricci scalar'' that we take along the x-axis in Fig.[\ref{plot_fR_solution}]. 
Recall, the Ricci scalar near the bounce gets the value
$R(t\rightarrow0) \simeq 12\alpha$ where the parameter $\alpha$
lies within $[3.208\times10^{-8}/\kappa^2 ,
3.300\times10^{-8}/\kappa^2]$ to make the primordial observable
quantities compatible with the Planck constraints. Thereby, the
Ricci scalar near the bounce becomes of the order $\sim
12\times10^{28}\mathrm{GeV}^2$, while in the present
epoch, the scalar curvature is generally considered to take
$10^{-84}\mathrm{GeV}^2$. Thus in order to correctly describe the
epochs from bounce to late time acceleration, we have to cover the
range of $R(t)$ from $\sim 12\times10^{28}\mathrm{GeV}^2$ to
$10^{-84}\mathrm{GeV}^2$. However in Fig.[\ref{plot_fR_solution}], we
consider a ``rescaled Ricci scalar'' given by $\bar{R} =
\frac{R}{\alpha}$ along the x-axis, which is dimensionless as
$\alpha$ has the mass dimension [+2]. Due to such rescaling, the
x-axis of the plot runs from the value $x_i = 10^{-112}$ (or we
may take $x_i = 0$) to $x_f = 12$. Moreover, the y-axis of
Fig.[\ref{plot_fR_solution}] is also rescaled by $\bar{F}(\bar{R})
= F(R)/\alpha$ so that the Einstein gravity can be described by a
straight line having slope of unity even in these rescaled
coordinates. The left part of the Fig. [\ref{plot_fR_solution}]
clearly demonstrates that the solution of $F(R)$ matches with the
Einstein gravity as the Ricci scalar approaches to the present
value, while the $F(R)$ seems to deviate from the usual Einstein
gravity, when the scalar curvature takes larger and larger values.
It is evident that near the bounce, $F'(R)$ is positive, which in
turn indicates the stability for the scalar and tensor
perturbations. This finding from the numerical solution, in fact,
resembles with the analytic results obtained in
Sec.[\ref{sec_estimation}]. The comparison of $F(R)$ gravity with
the Einstein one in the low-curvature regime is explicitly
demonstrated in the right part of Fig.[\ref{plot_Hubble
parameter}], which reveals that the form of $F(R)$ reconstructed
in the present context is indeed comparable with Einstein gravity
during the present epoch. Thus, we can safely argue that the
numerical solution of $F(R)$ obtained above indeed matches with
the analytic results determined near the bounce and at the dark
energy era.
\begin{figure}[!h]
\begin{center}
 \centering
 \includegraphics[width=3.5in,height=2.0in]{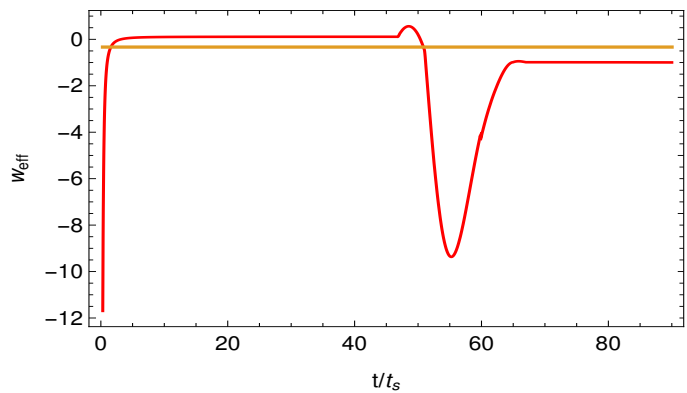}
 \caption{$w_{eff}$ (along y axis) vs. $\frac{t}{t_s}$ (along x axis)
 for $0 \lesssim t \leq 9\times10^{17}$ sec. The red curve represents the $w_{eff}$ for the present model while the yellow one is for the constant value
 $-\frac{1}{3}$. Thereby the successive crossing between the red and yellow curves depicts the transition of the Universe from acceleration to
 deceleration or vice-versa.}
 \label{plot_EoS}
\end{center}
\end{figure}
Using the form of $F(R)$ and the Hubble parameter from
Fig.[\ref{plot_fR_solution}] and [\ref{plot_Hubble parameter}]
respectively, we determine the remaining bit i.e the effective
equation of state (EoS) parameter defined by,
\begin{eqnarray}
 w_{eff} = \frac{\bigg[\frac{f(R)}{2} - \big(3H^2 + \dot{H}\big)f'(R) + 6\big(8H^2\dot{H} + 4\dot{H}^2 + 6H\ddot{H}
+ \dddot{H}\big)f''(R) + 36\big(4H\dot{H} + \ddot{H}\big)^2f'''(R)\bigg]}
{\bigg[-\frac{f(R)}{2} + 3\big(H^2 + \dot{H}\big)f'(R) - 18\big(4H^2\dot{H} + H\ddot{H}\big)f''(R)\bigg]}~~,
\nonumber
\end{eqnarray}
as shown in Fig. [\ref{plot_EoS}], where the x-axis is the
rescaled time coordinate i.e $\frac{t}{t_s}$ with $t_s = 10^{16}$
sec. Actually the red curve of the plot represents the $w_{eff}$
for the present model while the yellow one is for the constant
value $-\frac{1}{3}$ (we will keep the yellow graph to investigate
the accelerating or decelerating era of the Universe).
Fig.[\ref{plot_EoS}] clearly demonstrates that at the bounce i.e.
at $t = 0$, the EoS parameter diverges from the negative side,
however this is expected because at the bounce the Hubble
parameter itself becomes zero and in turn makes the $w_{eff} = -1 - \frac{2\dot{H}}{3H^2} \rightarrow -\infty$. 
Then immediately after the bounce,
$w_{eff}$ crosses the value $-\frac{1}{3}$ leading to a transition
from a bounce to a decelerating phase of the Universe, and during
the deceleration epoch, the EoS parameter remains almost constant
and near to zero indicating a quasi-matter dominated Universe
which continues till $t \simeq 5\times10^{17}$ sec, when the EoS
parameter again crosses the value $-\frac{1}{3}$ and thus the
Universe transits from a stage of deceleration to a stage of
acceleration continued until the  late-time era. Therefore, the
present model may provide an unified scenario of certain
cosmological epochs from bounce to late-time acceleration followed
by a quasi-matter dominated epoch in the intermediate regime. Moreover,
the EoS parameter during the dark energy era approaches to the
value $-0.995$, as evident from the Fig.[\ref{plot_EoS}]. This
late time value of the $w_{eff}$ is, however, in agreement with
\cite{Elizalde:2010ts}
where the authors considered the exponential $F(R)$ as a dark energy model, similar to the present case.

\section{Logarithmic corrected exponential $F(R)$ gravity as dark energy model:
A different $F(R)$ for unification of bounce and late time
acceleration}\label{sec_exponential-logFR}

 In this section, we
consider a different $F(R)$ dark energy model in comparison to the
one we considered in the previous sections, in particular, we
incorporate an additional logarithmic correction to the
exponential $F(R)$ gravity. Thus the form of $F(R)$ is given by
\cite{Odintsov:2018qug},
\begin{eqnarray}
 F(R) = R - 2\Lambda\bigg(1 - e^{-\frac{\beta R}{2\Lambda}}\bigg)\bigg[1 - \frac{\gamma R}{2\Lambda}~\log{\big(\frac{R}{4\Lambda}\big)}\bigg]
 \label{different DE}
\end{eqnarray}
where the correction over Einstein gravity is given by 
\begin{eqnarray}
F(R) - R = - 2\Lambda\bigg(1 - e^{-\frac{\beta R}{2\Lambda}}\bigg)\bigg[1 - \frac{\gamma R}{2\Lambda}~\log{\big(\frac{R}{4\Lambda}\big)}\bigg]~~.
\nonumber
\end{eqnarray}
It may be observed that such correction goes to zero at $R \rightarrow 0$ i.e in the low curvature regime. Now in order to avoid 
including an implicit cosmological constant in a $F(R)$ model or equivalently to avoid the fine tuning problem associated with the cosmological constant, 
the correction over the Einstein-Hilbert term of the $F(R)$ gravity requires to vanish in the low curvature limit i.e 
\begin{eqnarray}
\lim_{R\rightarrow 0}\big(F(R) - R\big) = 0~~.
\label{new_condition}
\end{eqnarray}
As just mentioned, this condition is satisfied for the considered $F(R)$ model in Eq.(\ref{different DE}) and thus 
this $F(R)$ model can produce the vacuum solution as the Minkowski spacetime. Thereby we may argue that the model (\ref{different DE}) is free 
from the problem of including an implicit cosmological constant, unlike to the scenario with the action given by 
$S = \int d^4x\sqrt{-g}\big[R - 2\Lambda\big]$. 
The condition (\ref{new_condition}) for avoiding a cosmological constant is also satisfied for the exponential $F(R)$ gravity 
as considered earlier in Eq.(\ref{rec_late_FR}) i.e 
without the logarithmic corrections. However as described in \cite{Odintsov:2018qug}, 
the presence of the logarithmic correction, modelled by the free parameter $\gamma$ by Eq.(\ref{different DE}), 
leads to a better fit (in respect to a viable dark energy model) 
than in absence of the logarithmic correction and also better than the $\Lambda\mathrm{CDM}$ model. Hence 
the inclusion of the logarithmic correction provides a test for this 
type of $F(R)$ theory and motivates to check how a deviation is allowed in comparison to the $\Lambda\mathrm{CDM}$ model. 
Note that the logarithmic correction introduces a new parameter
$\gamma$ i.e for $\gamma = 0$, the $F(R)$ model (\ref{different
DE}) becomes similar to the exponential model considered
earlier in Eq.(\ref{rec_late_FR}). This new $F(R)$ model with the
logarithmic term still satisfies the viability conditions (under
some conditions of the free parameter) and provides an extra term
in the action that evolutes smoothly along the cosmological
evolution (far from the pole obviously), as addressed in the Ref.
\cite{Odintsov:2018qug}. In particular, the logarithmic corrected
exponential $F(R)$ gravity comes as a viable dark energy model
with respect to the Sne-Ia+BAO+H(z)+CMB data for the parametric
choices : $\gamma = 0.0014^{+0.0025}_{-0.0014}$, $\Lambda =
1.2\times10^{-84}\mathrm{GeV}^2$ and $\beta = 4.71^{+\infty}_{-2.87}$
respectively. Such parametric regimes makes the $F(R)$ model free
from antigravity effects due to the condition $\gamma
\log{\big(\frac{R}{4\Lambda}\big)} < 1$ holds true. The authors of
\cite{Odintsov:2018qug} clearly investigated the possible effects
of the logarithmic term on the dark energy epoch in view of the
aforementioned observational data and as a result, they found that
the logarithmic correction modelled by the parameter $\gamma$
leads to better fit than in the absence of logarithmic correction
(i.e. the pure exponential $F(R)$ model), obviously at the price
of introducing a new parameter $\gamma$. Moreover the model
(\ref{different DE}) avoids the presence of large corrections on
the Newton's law as well as the appearance of large instabilities
at local systems, leading to a suitable model that recovers the
well known results of General Relativity at the appropriate
scales. In view of this new dark energy model, we propose the
following $F(R)$ gravity,
\begin{eqnarray}
 F(R) = R&-&2\Lambda\bigg(1 - e^{-\frac{\beta R}{2\Lambda}}\bigg)\bigg[1 - \frac{\gamma R}{2\Lambda}~\log{\big(\frac{R}{4\Lambda}\big)}\bigg]
 + \exp{\bigg\{1 - \cosh{\bigg(\frac{R}{R_b} + \frac{R_b}{R} - 2\bigg)}\bigg\}}\nonumber\\
 &\bigg[&\bigg(\frac{12\alpha D}{\sqrt{e}} - 1 - \gamma \log{\big(\frac{R}{4\Lambda}\big)}\bigg)R - D\sqrt{6\alpha\pi}~e^{-\frac{R}{24\alpha}}\big(R - 12\alpha\big)^{3/2}~
 Erfi\big[\frac{\sqrt{R - 12\alpha}}{2\sqrt{6\alpha}}\big]\bigg]
 \label{unifiedF(R)1}
\end{eqnarray}
which can merge the bounce and late-time acceleration as we will
show in the following, where the late time behavior is described
by the $F(R)$ of Eq.(\ref{different DE}) and the bouncing scenario
is still depicted by the $F_b(R)$ determined in
Eq.(\ref{rec_bounce sol2}). In the above expression, the $R_b$
denotes the scalar curvature at the bounce i.e $R_b = 12\alpha
\sim 10^{28}\mathrm{GeV}^2$ (see Sec. [\ref{sec_estimation}] for the
estimation of $\alpha$). We introduce an additional cosine
hyperbolic factor (let us call it ``fixing factor``) with the
second term of the right hand side of Eq.(\ref{unifiedF(R)1}) in
order to avoid the effects of the second term during the late-time
era. With the above $F(R)$, the demonstration of universe's evolution 
at different curvature regimes goes as
follows: (1) for $R \gg \Lambda$ when the Ricci scalar is given by
$R(t) = 12\alpha + 12\alpha^2t^2$ (i.e near the bounce), the
cosine hyperbolic term behaves as,
\begin{eqnarray}
 \cosh{\bigg(\frac{R}{R_b} + \frac{R_b}{R} - 2\bigg)} = \cosh{\bigg(\frac{\alpha^2t^4}{(1 + \alpha t^2)}\bigg)}
 = \frac{1}{2}\big[e^{\frac{\alpha^2t^4}{(1 + \alpha t^2)}} + e^{-\frac{\alpha^2t^4}{(1 + \alpha t^2)}}\big]
 &=&1 + \frac{\alpha^4t^8}{2(1 + \alpha t^2)^2} + .........\nonumber\\
 &=&1 + \mathcal{O}(t^8)
 \nonumber
\end{eqnarray}
and therefore the fixing factor can be expressed by $e^{-\mathcal{O}(t^8)} =
1 - \mathcal{O}(t^8)$. However, recall that the near bounce quantities are
determined up to $\mathcal{O}(t^2)$ and thus the factor $e^{-\mathcal{O}(t^8)}$ can be
approximated to unity near $R \simeq R_b$. As a result along with
the fact $R_b \gg \Lambda$, the $F(R)$ of Eq.(\ref{unifiedF(R)1})
turns out to be,
\begin{eqnarray}
F(R) = \frac{12\alpha D}{\sqrt{e}}R - D\sqrt{6\alpha\pi}~e^{-\frac{R}{24\alpha}}\big(R - 12\alpha\big)^{3/2}
Erfi\bigg[\frac{\sqrt{R - 12\alpha}}{2\sqrt{6\alpha}}\bigg] = F_b(R)
\label{new1}
\end{eqnarray}
in the regime $R \simeq R_b$, which realizes a bouncing
Universe as discussed in Sec. [\ref{sec_rec_bounce}]. However in
order to ensure a bounce, it is also necessary to check $F'(R)$
and $F''(R)$ as they are connected to the issues of energy conditions. 
Being the evolution of the fixing factor as
$e^{-\mathcal{O}(t^8)} = 1 - \mathcal{O}(t^8)$ and since $\frac{dR}{dt}$ is
proportional to $t$ around the bounce, one can immediately write
$F'(R) = F_b'(R)$ and $F''(R) = F_b''(R)$ up to $\mathcal{O}(t^2)$ near $R
\simeq R_b$. Thereby, the $F(R)$ we propose in
Eq.(\ref{unifiedF(R)1}) as well as its first and second
derivatives match with that of $F_b(R)$ up to $\mathcal{O}(t^2)$, which
clearly indicates a bouncing Universe for $R \simeq R_b$. Here it
may be mentioned that a different fixing factor like
$e^{-\big(\frac{R}{R_b} + \frac{R_b}{R} - 2\big)}$ behaves as $1 +
\mathcal{O}(t^4)$ near the regime $R \simeq R_b$, unlike to our considered
fixing factor where the leading order is proportional to $t^8$ due
to the additional cosine hyperbolic term. However in effect of
fourth order term (i.e $t^4$) being the leading order one, the
fixing factor $e^{-\big(\frac{R}{R_b} + \frac{R_b}{R} - 2\big)}$
makes the $F'(R)$ different in comparison to $F_b'(R)$ even up to
$\mathcal{O}(t^2)$, which in turn may violate the bounce at $R \simeq R_b$. 
In view of these arguments, we stick to the ``fixing factor'' as proposed 
in Eq.(\ref{unifiedF(R)1}) rather than $e^{-\big(\frac{R}{R_b} + \frac{R_b}{R} - 2\big)}$.
On other hand, (2) for $R \ll R_b$ i.e in the low curvature regime
(or equivalently $R \sim \Lambda$), $\frac{R_b}{R}$ becomes very
much larger in comparison to $\frac{R}{R_b}$ and as a consequence
the factor $\exp{\bigg\{1 - \cosh{\bigg(\frac{R}{R_b} +
\frac{R_b}{R} - 2\bigg)}\bigg\}}$ can be approximated to
$\exp{\big[-e^{R_b/R}\big]}$ which further tends to zero in the
regime $\frac{R}{R_b} \ll 1$. Thereby, in the low curvature
regime, the form of $F(R)$ can be written as $F(R) = R -
2\Lambda\bigg(1 - e^{-\frac{\beta R}{2\Lambda}}\bigg)\bigg[1 -
\frac{\gamma
R}{2\Lambda}~\log{\big(\frac{R}{4\Lambda}\big)}\bigg]$ leading to
a viable dark energy dominated era in respect to
Sne-Ia+BAO+H(z)+CMB observations for a suitable parametric spaces, as
mentioned earlier. Moreover the $F(R)$ model (\ref{unifiedF(R)1}) satisfies $\lim_{R\rightarrow 0}\big(F(R) - R\big) = 0$ 
i.e the correction over the 
Einstein-Hilbert term vanishes in the limit $R \rightarrow 0$. This indicates that the $F(R)$ of Eq.(\ref{unifiedF(R)1}) is able to provide 
the Minkowski solution in vacuum case and thus is free from the problem 
of including an implicit cosmological constant. Thus, the $F(R)$ proposed in Eq.
(\ref{unifiedF(R)1}) appropriately provides a non-singular bounce
and a late acceleration near $R \simeq R_b$ and during $R \sim
\Lambda$ respectively, and hence is able to merge a
bounce with dark energy (DE) epoch. The deceleration stage in
between the bounce and the dark energy epochs can be obtained by
the numerical interpolation, similarly as we did for the earlier
case in Sec.[\ref{sec_unification}].\\

At this stage it deserves mentioning that we have taken the bottom-up approach to reconstruct the $F(R)$ gravity in the present context, 
in which we attempt to figure out the functional form of $F(R)$ by demanding that it should
explain the observational results. However, various terms in the $F(R)$ of Eq.(\ref{unifiedF(R)1}) may be thought to have a fundamental origin, like - 
near the bouncing regime, the $F(R)$ (denoted by $F_b(R)$) is given by Eq.(\ref{new1}). Thereby if we expand the Taylor series of the function 
$Erfi\bigg[\frac{\sqrt{R - 12\alpha}}{2\sqrt{6\alpha}}\bigg]$ (present in Eq.(\ref{new1})) around the bouncing point 
i.e around $R = 12\alpha$ and keeping the leading order term, then the near-bounce expression of $F(R)$ will contain linear and quadratic 
power in curvature $R$. Such quadratic correction in the Ricci scalar plays a role for renormalizability of General Relativity (GR) in quantum gravity. 
On other hand, during the late time, the $F(R)$ behaves as of Eq.(\ref{different DE}). The logarithmic correction present in the late time form of 
$F(R)$ may be thought as one-loop effects of quantum gravity. Despite these arguments, it is true that due to the complicated nature 
of the full $F(R)$ presented in Eq.(\ref{unifiedF(R)1}), it is hard to realize some fundamental origin of this full form of $F(R)$. Of course, 
the complete gravitational action should be defined 
by a fundamental theory, which, however, remains to be the open problem of modern high-energy physics. In the absence of 
fundamental quantum gravity, the modified gravity approach in this work is a phenomenological one that is constructed by 
complying with observational data.\\

Before concluding, we would like to mention that the present scenario merges certain cosmological epochs of the universe, 
in particular, from a non-singular bounce to a matter dominated epoch
and from the matter dominated to a late time accelerating epoch; 
i.e the model is similar to a matter bounce model which is also compatible to a late dark energy phase of the cosmic evolution as well. 
However this is not the full evolution history of the universe mainly due to the absence of the radiation era. 
Thereby in order to unify the entire evolutionary epochs of the universe in the context of bouncing cosmology, one should show the unification 
of an early bounce with the radiation domination, followed by the photon decoupling, followed by matter domination, all
the way to the late-time dominance of dark energy stage. The unification for the full evolution of the universe is a longstanding problem 
in cosmology and we hope that our present paper may enlighten some part(s) of this bigger problem.

\section{Conclusion}\label{sec_conclusion}

In the present work, we proposed a cosmological model in the context of $F(R)$ gravity, which merges a non-singular bounce to a matter dominated epoch
and from the matter dominated to a late time accelerating epoch; i.e the model 
is similar to a generalized matter bounce model which is also compatible to a late dark energy dominant phase of the cosmic evolution. 
Using the reconstruction schemes,
we reconstructed the form of $F(R)$ near the bounce and at the
late-time era respectively. However, the reconstruction techniques
applied in these two eras are slightly different, in particular:
near the bouncing regime, we first considered a scale factor
(suitable for bounce) and then determined the form of $F(R)$ which
can realize such bouncing scale factor by using the gravitational
equation of motion, while on other hand in the case of late times, we have
used a reverse reconstruction procedure in comparison to that of the earlier one i.e. we
started with a form of $F(R)$ (rather than a scale factor)
viable for dark energy model, namely the exponential $F(R)$
gravity, and then reconstructed the corresponding Hubble parameter
from the gravitational equations of motion. During such
reconstructions, we got early and late stage model parameters
which have been estimated from various observational constraints. The
early stage parameters have been obtained from the primordial
perturbations and we confronted the theoretical results of the
observable quantities like the scalar spectral index, the
tensor-to-scalar ratio with the latest Planck 2018 constraints. On
other hand, the late stage parameters are estimated by
investigating the viability of the exponential $F(R)$ gravity as a
dark energy model with respect to the Sne-Ia+BAO+H(z)+CMB data.
Due to a late accelerating phase, the Hubble radius decreases
monotonically at late times and asymptotically goes to zero, which
in turn leads to the generation of the primordial perturbation 
near the bounce (because at that time the relevant perturbation
modes are within the horizon), unlike to the usual matter bounce
scenario where the perturbation modes generate deeply in the
contracting regime far away from the bouncing point time. Thus
we have performed the perturbations near the bounce and as a
result the primordial observable quantities like the spectral
index for curvature perturbation, and the tensor-to-scalar ratio,
are found to be simultaneously compatible with the latest Planck
2018 observations. Moreover, the scalar and tensor perturbations
are stable as the condition $F'(R) > 0$ holds in the present
context. With the early bounce and the late accelerating phase,
we provided a complete evolution of the Hubble radius by
considering the intermediate region to be a matter dominated epoch
and the transitions from the bounce to matter dominated and from
the matter dominated to the DE era have been obtained by the
method of numerical interpolation, where the estimated values of
the model parameters are incorporated. The evolution of the Hubble
radius provides the universe's evolution for a considerable range of cosmic time, in particular, it merges a non-singular 
bounce to a matter dominated epoch, followed by a late time accelerating stage. 
With this interpolating Hubble radius, we have numerically solved the $F(R)$ gravitational
equation to determine the form of $F(R)$ for a wide range of
cosmic time, which clearly depicts that the $F(R)$ matches with
the Einstein gravity in the low curvature regime, while it
deviates from the usual Einstein gravity as the scalar curvature
acquires larger and larger values. Correspondingly, the Hubble
parameter and the effective EoS parameter of the Universe have
been determined. As a result the EoS parameter is found to
successively cross the value $-\frac{1}{3}$ twice indicating the
transition of the Universe from bounce to a quasi-matter dominated phase and from the quasi-matter
dominated to a late-time accelerating stage respectively. Moreover, the
EoS parameter during the dark energy era approaches to the value
$-0.995$. Finally, in Sec. [\ref{sec_exponential-logFR}], we have
proposed a different $F(R)$ gravity form which is able to merge a
non-singular bounce with a dark energy epoch, however the DE epoch
is described by a logarithmic corrected exponential $F(R)$
gravity, unlike to the earlier case where the dark energy model is
depicted by the usual exponential $F(R)$ gravity i.e. without the
logarithmic corrections. Similar to the exponential $F(R)$ model,
the logarithmic generalized exponential $F(R)$ gravity is also
known to provide a viable dark energy model with respect to
Sne-Ia+BAO+H(z)+CMB observations, in fact, the presence of
logarithmic corrections leads to a better fitted model than the
usual exponential $F(R)$ model.

\begin{acknowledgments}
This work is supported by MINECO (Spain), FIS2016-76363-P (S.D.O).
\end{acknowledgments}

\end{document}